\documentclass[a4paper,fleqn,usenatbib]{mnras}
\pdfoutput=1
\usepackage{newtxtext,newtxmath}
\usepackage[T1]{fontenc}
\usepackage{ae,aecompl}

\usepackage{graphicx}
\usepackage{amsmath,amssymb}
\usepackage{gensymb}

\newcommand{\lya}{Ly$\alpha$}

\newcommand{\ciii}{\hbox{C\,{\sc iii}}}
\newcommand{\civ}{\hbox{C\,{\sc iv}}}
\newcommand{\oiii}{\hbox{O\,{\sc iii}}}
\newcommand{\oii}{\hbox{O\,{\sc ii}}}

\newcommand{\nv}{\hbox{N\,{\sc v}}}

\newcommand{\siII}{\hbox{Si\,{\sc ii}}}

\newcommand{\hii}{\hbox{H\,{\sc ii}}}

\newcommand{\heii}{\hbox{He\,{\sc ii}}}

\newcommand{\wciv}{\ensuremath{\rm W_{C{\scriptscriptstyle IV},0}}}
\newcommand{\whb}{\ensuremath{\rm W_{H\beta,0}}}

\newcommand{\unit}[1]{\ensuremath{\, \mathrm{#1}}}

\newcommand{\hst}{\textit{HST}}
\newcommand{\hstcos}{\textit{HST}/COS}
\newcommand{\fos}{\textit{HST}/FOS}
\newcommand{\jwst}{\textit{JWST}}

\newcommand{\ott}{\ensuremath{\mathrm{O}_{32}}}
\newcommand{\rtt}{\ensuremath{\mathrm{R}_{23}}}

\newcommand{\cloudy}{\textsc{cloudy}}

\newcommand{\beagle}{\textsc{beagle}}
\newcommand{\tauV}{\hbox{$\hat{\tau}_V$}}

\defcitealias{Senchyna2017}{S17}

\title[Extremely metal-poor galaxies in the UV]{Extremely metal-poor galaxies with \hstcos{}: laboratories for models of low-metallicity massive stars and high-redshift galaxies}
\author[P. Senchyna et al.]{
    Peter Senchyna$^{1}$\footnotemark[1],
    Daniel P. Stark$^{1}$,
    Jacopo Chevallard$^{2}$,
    St\'{e}phane Charlot$^{3}$, \newauthor
    Tucker Jones$^{4,5}$ and
    Alba Vidal-Garc\'ia$^{3,\, 6}$\\
    \vspace{0.1in}\\
    $^{1}$ Steward Observatory, University of Arizona, 933 N Cherry Ave, Tucson, AZ 85721 USA \\
    $^{2}$ Scientific Support Office, Directorate of Science and Robotic Exploration, ESA/ESTEC, Keplerlaan 1, 2201 AZ Noordwijk, The Netherlands \\
    $^{3}$ Sorbonne Universit\'{e}, CNRS, UMR7095, Institut d'Astrophysique de Paris, F-75014, Paris, France \\
    $^{4}$ Department of Physics, University of California Davis, 1 Shields Avenue, Davis, CA 95616, USA \\
    $^{5}$ Institute of Astronomy, University of Hawaii, 2680 Woodlawn Drive, Honolulu, HI 96822, USA \\
    $^{6}$ Laboratoire de Physique de l'ENS, ENS, Universit\'{e} PSL, CNRS, Sorbonne Universit\'{e}, Universit\'{e} Paris-Diderot, Paris, France \\
}

\date{Accepted XXX. Received YYY; in original form ZZZ}

\pubyear{2019}

\begin{document}
\label{firstpage}
\pagerange{\pageref{firstpage}--\pageref{lastpage}}
\maketitle

\begin{abstract}
    Ultraviolet (UV) observations of local star-forming galaxies have begun to establish an empirical baseline for interpreting the rest-UV spectra of reionization-era galaxies.
    However, existing high-ionization emission line measurements at $z>6$ ($\wciv{} \gtrsim 20$ \AA{}) are uniformly stronger than observed locally ($\wciv{} \lesssim 2$ \AA{}), likely due to the relatively high metallicities ($Z/Z_\odot > 0.1$) typically probed by UV surveys of nearby galaxies.
    We present new \hstcos{} spectra of six nearby ($z<0.01$) extremely metal-poor galaxies (XMPs, $Z/Z_\odot \lesssim 0.1$) targeted to address this limitation and provide constraints on the highly-uncertain ionizing spectra powered by low-metallicity massive stars.
    Our data reveal a range of spectral features, including one of the most prominent nebular \civ{} doublets yet observed in local star-forming systems and strong \heii{} emission.
    Using all published UV observations of local XMPs to-date, we find that nebular \civ{} emission is ubiquitous in very high specific star formation rate systems at low metallicity, but still find equivalent widths smaller than those measured in individual lensed systems at $z>6$.
    Our moderate-resolution \hstcos{} data allow us to conduct an analysis of the stellar winds in a local nebular \civ{} emitter, which suggests that some of the tension with $z>6$ data may be due to existing local samples not yet probing sufficiently high $\mathrm{\alpha/Fe}$ abundance ratios.
    Our results indicate that \civ{} emission can play a crucial role in the \jwst{} and ELT era by acting as an accessible signpost of very low metallicity ($Z/Z_\odot < 0.1$) massive stars in assembling reionization-era systems.
\end{abstract}


\footnotetext[1]{E-mail: senchp@email.arizona.edu}

\newpage
\clearpage

\section{Introduction}

The first spectroscopic detections of emission lines other than \lya{} in reionization-era galaxies have proved to be surprisingly challenging to interpret.
Prominent emission in \ciii{}], \civ{}, \heii{}, and \nv{} detected at $z>6$ belies the presence of very hard ionizing radiation fields at these early times, and is essentially without precedence in lower-redshift star-forming galaxy samples \citep[e.g.][]{Stark2015,Stark2015a,Sobral2015,Stark2017,Mainali2017,Laporte2017,Mainali2018,Dors2018,Shibuya2018,Sobral2019}.
Preliminary modeling of this emission suggests that the observed \ciii{}], \civ{}, and \heii{} may be the signature of very low-metallicity massive stars.
However, this interpretation is presently extremely uncertain due to limitations in our understanding of the ionizing spectra of metal-poor massive stars \citep[e.g.][]{Crowther2006,Levesque2012,Goetberg2018,Stanway2018}.

The stellar population synthesis models used to interpret high-redshift galaxy observations are anchored by observations in the local Universe.
Spectra of individual massive stars have been used to calibrate the effective temperatures and stellar winds of the massive O and B stars which dominate the ionizing continuum in young galaxies \citep[e.g.][]{Kudritzki1987,Massey2005,Crowther2006,Tramper2011,Garcia2014,Crowther2016}.
Unfortunately, this is only possible with current facilities inside the Local Group where individual stars can be reliably resolved, which limits this work to metallicities above $Z/Z_\odot \simeq 0.1$ \citep[e.g.][]{Vink2001,Bouret2015,Camacho2016}.
As a result, effective temperature and wind scaling relations have not been directly calibrated for massive stars at lower metallicities, and predictions of the ionizing spectra these stars power are essentially entirely theoretical \citep[e.g.][]{Lejeune1997,Graefener2002,Smith2002,Lanz2003}.

Nearby but unresolved star-forming galaxies outside the Local Group provide laboratories in which to study populations of massive stars at ages and metallicities relevant to reionization-era galaxy models.
Most studies with early UV spectroscopic instruments focused on relatively bright, massive, relatively high-metallicity star-forming systems \citep[typical metallicity $12+\log\mathrm{O/H} \simeq 8.2$:][]{Leitherer2011}; but the harder stellar ionizing spectra and higher gas temperatures expected in low metallicity \hii{} regions are likely conducive to stronger high-ionization nebular emission.
The Cosmic Origins Spectrograph onboard the Hubble Space Telescope (\hstcos{}) has revolutionized this work, providing the sensitivity and resolution necessary to investigate the ultraviolet high-ionization nebular and wind features powered by the youngest stars in relatively faint low-metallicity galaxies.
In \hst{} Cycle 23, we obtained moderate-resolution G160M and G185M COS spectra of 10 star-forming regions characterized by very large specific star formation rates (sSFRs $\sim 10^2$ \unit{Gyr^{-1}}) and nebular or stellar \heii{} emission in the optical, targeting \civ{} $\lambda \lambda 1548, 1550$, \heii{} $\lambda 1640$, \oiii{}] $\lambda \lambda 1661, 1666$, and \ciii{}] $\lambda \lambda 1907,1909$ \citep[][hereafter \citetalias{Senchyna2017}]{Senchyna2017}.
The targets in this initial study spanned metallicities $7.8\leq 12+\log\mathrm{O/H} \leq 8.5$ (i.e.\ $0.1\lesssim Z/Z_\odot \lesssim 0.6$), and reveal a significant variation in nebular emission over this range.
Nebular \civ{} emission and the highest equivalent width \ciii{}] both appear in the most metal-poor galaxies in the sample at $12+\log\mathrm{O/H}<8.0$, suggestive of a rapid hardening of the ionizing spectrum below $Z/Z_\odot < 0.2$ \citep[see also][]{Chevallard2018}.
While this study confirmed that stars can power nebular \civ{} emission, the equivalent width of this doublet in the \citetalias{Senchyna2017} sample still falls an order of magnitude short of the $\gtrsim 20$ \AA{} emission observed in the reionization era.
Due to the limited observed sample size, the empirical question of whether the massive stars encountered in extremely metal-poor galaxies (XMPs) at $12+\log\mathrm{O/H}<7.7$ ($Z/Z_\odot < 0.1$) are capable of powering nebular emission similar to that observed at $z>6$ remains unanswered.

To address this paucity of data, in \hst{} Cycle 24 we obtained \hstcos{} observations of six star-forming XMPs with the moderate-resolution grisms G160M and G185M necessary to disentangle nebular emission from stellar and interstellar absorption in \civ{} (GO: 14679, PI: Stark).
In addition, we gather all archival ultraviolet observations of XMPs which cover \ciii{}] and \civ{}.
In this paper, we provide the first systematic analysis of these nebular lines and massive stellar wind features in all local XMPs with ultraviolet constraints, directly assessing the range of nebular emission powered by $Z/Z_\odot<0.1$ stellar populations.
We outline the data and measurements in Section~\ref{sec:data}, present our new results in Section~\ref{sec:results}, discuss implications for stellar population modeling in Section~\ref{sec:discuss}, and conclude in Section~\ref{sec:summary}.

For comparison with solar metallicity, we assume a solar gas-phase oxygen abundance of $12+\log_{10}\left([\text{O/H}]_\odot\right)  = 8.69$ \citep{Asplund2009} unless otherwise noted.
For distance calculations and related quantities, we adopt a flat cosmology with $H_0 = 70$ \unit{km \, s^{-1} \, Mpc^{-1}}.
All equivalent widths are measured in the rest-frame and we choose to associate emission with positive values ($W_0>0$).

\section{Data and Analysis}
\label{sec:data}

\subsection{Archival Context}
\label{sec:uvarchive}

Unfortunately, XMPs have rarely been included in ultraviolet spectroscopic campaigns up to this point.
Here we assess the status of the literature on XMPs in the UV to provide context for our sample and for inclusion in our analysis later in this paper.
Measurements are presented in the appendix in Table~\ref{tab:archival_data}.
The legacy spectroscopic atlas assembled in \citet{Leitherer2011} summarized the state of ultraviolet spectra prior to the installation of \hstcos{}, consisting of essentially all useful spectra of star-forming galaxies obtained with the Faint Object Spectrograph (FOS) and Goddard High Resolution Spectrograph (GHRS).
The earlier International Ultraviolet Explorer telescope (IUE) only allowed $R\ll 1000$ spectra and achieved relatively low S/N, making study of \ciii{}] challenging and \civ{} essentially impossible \citep[e.g.][]{Dufour1988}.
The \citet{Leitherer2011} sample contained spectra of regions in four XMPs: SBS 0335-052, I Zw 18, SBS 1415+437, and Tol 1214-277.
However, while \ciii{}] constraints are available for these systems, the wavelength settings of these observations miss \civ{} $\lambda \lambda 1548, 1550$ in all four cases.

Despite its improved sensitivity and spectral resolution, \hstcos{} has not yet been used extensively to study XMPs.
Among published \hstcos{} observations with coverage in the 1400--1700 \AA{} range, we find a small number of local star-forming galaxies with gas-phase metallicities $12+\log\mathrm{O/H}<7.7$.
Subregions of the canonical very nearby XMPs I Zw 18, SBS 0335-052, and DDO 68 have been targeted at moderate-resolution with \hstcos{}, but analysis of the nebular emission in these systems will be presented in Wofford et al.\ (in-preparation).
A total of 15 galaxies with gas-phase oxygen abundances in the XMP regime were observed at lower resolution with G140L ($R\sim 2000$) and presented by \citet{Berg2016,Berg2019}, including coverage of \civ{}.
We measure flux in or upper limits on the combined \civ{} emission in these spectra in the same way as for nebular lines in our spectra (see Section~\ref{sec:optspec}).
Measurements of \civ{} emission in these observations with the COS/G140L grating may be underestimated due to blending of interstellar absorption with nebular emission at this lower spectral resolution, but since we cannot directly remove this we proceed with the line measurements as-is (this absorption is likely small; see Figure~\ref{fig:majorlines_comp}).
To summarize, we are able to place constraints on \civ{} emission in fifteen XMPs with G140L spectra from the literature, including detections at equivalent widths ranging from 1--11 \AA{} (Table~\ref{tab:archival_data}).

\subsection{XMP Sample Selection for \hstcos{}}
\label{sec:sample}

Motivated by the lack of moderate-resolution UV spectra available at these lowest metallicities, we selected six new XMPs for study with the Cosmic Origins Spectrograph in HST Cycle 24.
Our target selection was guided by several considerations.
First, we focused on redshifts $z<0.03$, where the \ciii{}] $\lambda\lambda 1907,1909$ doublet is still accessible to the G185M grism.
Second, we aimed to span at least 0.3 dex below $12+\log\mathrm{O/H}<7.7$ to investigate metallicity dependence in this regime.
Finally, we required the targets to have a compact star-forming region bright enough in the UV to be studied in a single orbit per grism.
In particular, we selected systems with GALEX NUV magnitudes $i\lesssim 18$, which yield ETC-predicted signal-to-noise ratios $>3$ per resolution element in the G185M continuum at 1900 \AA{} in one orbit (assuming a flat continuum in $F_\nu$).
We selected 6 XMPs without existing ultraviolet spectra satisfying these requirements from the archival sample of XMPs assembled in \citet{Morales-Luis2011}.
The GALEX NUV magnitudes (ranging from 16.3 to 17.9) of the 6 selected targets span the brightest quartile of the \citet{Morales-Luis2011} archival catalog.
Their basic properties are summarized in Table~\ref{tab:basicprop}.

\begin{table*}
    \centering
	\caption{
        Basic data and \hstcos{} exposure times for the 6 XMPs studied in this work. The optical $u$ and $i$ band measurements are from SDSS imaging except for J0405-3648 (marked by a \textdagger), where DES $i$-band imaging is used (marked by a \textdagger).
    }
	\label{tab:basicprop}
  \begin{tabular}{lccccccccc}
  \hline
  Name  & RA & Dec & z & Distance & GALEX NUV & $u$ & $i$ & $i$ & NUV/G185M, FUV/G160M\\
   & (J2000) & (J2000) &  & (Mpc) & AB mag & fiber & fiber & total & exposure (s)\\
  \hline
  HS1442+4250 & 14:44:11.46 & 42:37:35.6 & 0.0023 & 11 & 16.3 & 18.7 & 19.0 & 15.7 & 2204, 2673 \\
  J0405-3648 & 4:05:20.46 & -36:48:59.1 & 0.0028 & 15 & 17.0 & --& 19.3 \textdagger & 16.3 \textdagger & 4328, 2665 \\
  J0940+2935 & 9:40:12.87 & 29:35:30.2 & 0.0024 & 8 & 17.0 & 18.9 & 18.8 & 16.0 & 2124, 2624 \\
  J1119+5130 & 11:19:34.37 & 51:30:12.0 & 0.0045 & 22 & 17.2 & 18.5 & 18.2 & 17.1 & 2300, 2801 \\
  SBSG1129+576 & 11:32:02.64 & 57:22:36.4 & 0.0050 & 25 & 17.3 & 19.6 & 19.3 & 16.3 & 2352, 2852 \\
  UM133 & 1:44:41.37 & 4:53:25.3 & 0.0053 & 29 & 17.9 & 18.7 & 19.0 & 15.6 & 2108, 2605 \\
  \hline
  \end{tabular}

\end{table*}

Distance estimates are necessary to compute total stellar masses and star formation rates.
Because these systems reside at redshifts $cz\lesssim 1500 \mathrm{km/s}$, peculiar velocities can significantly affect their apparent position in a smooth Hubble flow.
We uniformly compute luminosity distances using the local velocity flow model presented by \citet{Tonry2000}, which provides coarse-grained corrections for the peculiar velocities of galaxies in nearby clusters.
We assume $H_0=70 \unit{km}\unit{s}^{-1}\unit{Mpc}^{-1}$ \citep{Hinshaw2013} and redshifts measured from strong nebular lines in the optical spectra (see Section~\ref{sec:optspec}).
This procedure yields distances ranging from 8--29 Mpc for the newly-targeted XMPs (Table~\ref{tab:basicprop}).
At these distances, the 2.5\arcsec{} COS aperture corresponds to physical scales of 100--400 pc.
We will discuss the broadband imaging in the following subsection \ref{sec:imaging}, and proceed to describe the optical and UV spectra in Sections~\ref{sec:optspec} and \ref{sec:uvspec}, respectively.

\subsection{Imaging}
\label{sec:imaging}

\begin{figure*}
    \includegraphics[width=0.7\textwidth]{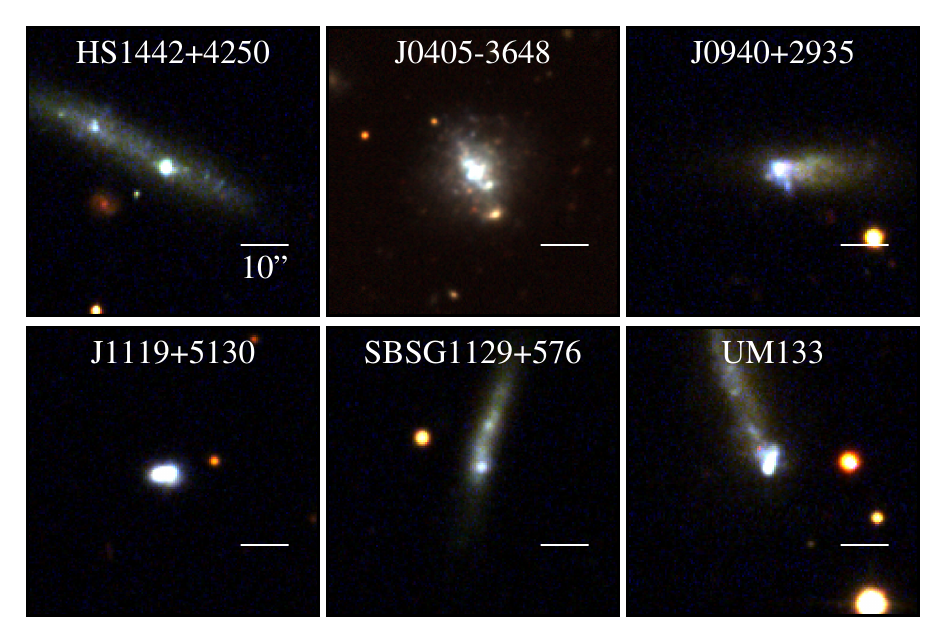}
    \caption{
        Optical imaging centered on our targets, ordered alphabetically.
        We plot DES DR1 $gri$ imaging for J0405-3648 and SDSS $ugr$ montages for the others.
        The objects under study are the bright compact star-forming regions centered in each image and targeted with the 2.5\arcsec{} \hstcos{} aperture.
    }
    \label{fig:cutouts}
\end{figure*}

Broadband optical imaging provides important information about the stellar populations underlying the ultraviolet spectra.
We plot optical mosaic images of our targets in Figure~\ref{fig:cutouts}.
Sloan Digital Sky Survey (SDSS) $ugriz$ imaging covers five of the six galaxies studied in this paper, while J0405-3648 was imaged by DECam in $grizY$ as part of the first data release of the Dark Energy Survey \citep[DES:][]{Abbott2018}.
Since this paper is concerned primarily with optical and ultraviolet spectra probing regions on the scale of $\lesssim 3$\arcsec{}, we fit SDSS aperture photometry measured within a 1.5\arcsec{} radius centered on the \hstcos{} targets.
This photometric aperture is standard for SDSS fiber analysis, and matches the size of the 1.25\arcsec{} radius \hstcos{} aperture after convolving with the typical SDSS seeing (1.35\arcsec{} FWHM in the $r$-band).

We use the SED fitting software \beagle{}\footnote{Version 0.19.9; \url{http://www.jacopochevallard.org/beagle/}} \citep{Chevallard2016} to derive constraints on the recent star formation in these objects.
In particular, we adopt the full \citet{Gutkin2016} model grid which combines the latest version of the \citet{Bruzual2003} stellar population synthesis results with the photoionization code \cloudy{} \citep{Ferland2013} to self-consistently fit the combination of stellar continuum and nebular emission underlying the broadband photometry.
We fit the circular 3\arcsec{} aperture SDSS/DES photometry in all available optical bands alongside the equivalent width of H$\beta$ emission measured from the optical spectra (Section~\ref{sec:optspec}) to explicitly constrain nebular band contamination.
We assume a constant star formation history\footnote{As discussed in \citep{Senchyna2019}, the assumption of a constant star formation history minimizes the number of free parameters being fit and allows for consistent comparison with SED fitting conducted at $z>6$, but  will yield systematically lower stellar mass estimates than fits assuming a declining or multiple-component star-formation history.} and allow the star formation rate, stellar mass, metallicity (gas-phase and stellar, which we assume are identical here), gas ionization parameter $\log U$, and effective attenuation \tauV{} to vary.
For this photometric fitting, we adopt the 2-component attenuation model presented by \citet{Charlot2000}.
The redshift and distance is fixed as measured in Table~\ref{tab:basicprop}.
In addition to the aperture magnitudes fit here, we also measure total flux of each galaxy in the $i$-band from a segmented image \citep[produced using {\texttt photutils}:][]{Bradley2018} and present these measurements in Table~\ref{tab:basicprop}.
This total flux ranges from 3--24 times that measured in the circular aperture, so assuming a constant mass-to-light ratio the total stellar masses may be larger by up to 1.4 dex.

The resulting constraints on the stellar mass, star formation rate, and attenuation are displayed in Table~\ref{tab:sedres}.
These fits indicate that the stellar populations within 3\arcsec{} are low in stellar mass ($4.9 < \log_{10}(M/M_\odot) < 6.4$) and attenuation ($\tauV \leq 0.12$), and forming stars at a range of rates: with the specific star formation rate $\mathrm{sSFR} = \mathrm{SFR} / M_\odot$ ranging from 2.1--18 $\mathrm{Gyr^{-1}}$.
These masses are comparable to those of the higher-metallicity galaxies presented in \citetalias{Senchyna2017}, but the sSFR values are lower than the 100 $\mathrm{Gyr^{-1}}$ which characterize those particularly extreme star-forming regions.
Correspondingly, we infer significantly older ages as well (in this constant star formation history framework, simply the inverse of the sSFR), ranging from 57--471 Myr compared to the median 10 Myr derived in \citetalias{Senchyna2017}.
This implies that we are viewing stellar populations with an older characteristic age in these XMPs.
These systems are lower in mass (as measured assuming a constant star formation history) than galaxies at $z>6$ with rest-frame UV metal line detections \citep[$10^9$--$10^{10}$ $M_\odot$, e.g.][]{Stark2015,Stark2015a,Mainali2017}, but the measured sSFR values extend into the range of values inferred for the general LBG population at $z\sim$6--8 \citep[e.g.][]{Labbe2013,Smit2014,Salmon2015,Faisst2016,Williams2018}.

\begin{table}
	\centering
	\caption{
        Results from broadband SED fitting assuming a constant star formation history with \beagle{}.
         These fits reveal relatively low total mass in stars within the spectroscopic aperture, $\log_{10}(M/M_\odot) \sim 4.9$ -- $6.4$), with specific star formation rates ranging from 2--18 $\mathrm{Gyr^{-1}}$.
        Note that the SFR indicated here is measured over the last 10 Myr.
    }
    \label{tab:sedres}
    \begin{tabular}{lccc}
        \hline
        Name & $\log_{10}(M/M_{\odot})$ & $\log_{10}(\mathrm{SFR}/(M_{\odot}/\mathrm{yr}))$ & sSFR ($\mathrm{Gyr^{-1}}$) \\
        \hline
        HS1442+4250 & $5.2 \pm 0.2$ & $-2.59 \pm 0.10$ & $17.5\pm8.2$\\
        J0405-3648 & $5.3 \pm 0.2$ & $-3.01 \pm 0.18$ & $4.6\pm3.4$\\
        J0940+2935 & $4.9 \pm 0.2$ & $-2.96 \pm 0.14$ & $12.6\pm6.6$\\
        J1119+5130 & $6.4 \pm 0.2$ & $-2.26 \pm 0.13$ & $2.3\pm1.1$\\
        SBSG1129+576 & $6.1 \pm 0.2$ & $-2.53 \pm 0.06$ & $2.1\pm1.0$\\
        UM133 & $5.9 \pm 0.2$ & $-1.96 \pm 0.09$ & $14.5\pm6.3$\\
        \hline
        \end{tabular}
\end{table}

\subsection{Optical Spectra}
\label{sec:optspec}

A uniform set of publicly-available optical spectra were not initially available for these XMPs, so we pursued observations with Blue Channel on the 6.5m MMT and the Echellete Spectrograph and Imager \citep[ESI,][]{Sheinis2002} on Keck II.
The former provides access to the strong lines including the auroral [\oiii{}] $\lambda 4363$ line and critically the (blended) [\oii{}] $\lambda\lambda 3727, 3729$ doublet missed by ESI.
ESI provides a high-resolution view of line complexes such as that near \heii{} $\lambda 4686$ which would otherwise be intractable to study with a lower resolution grating.

We used MMT Blue Channel to observe the five northern targets on January 25, 2017.
The systems were observed at airmasses $\leq 1.5$, and WFS measurements indicated seeing was steady near 1.3\arcsec{} throughout the night.
The observations are summarized in Table~\ref{tab:mmt_obs}.
The data were taken with the 300 $\mathrm{lines/mm}$ grating, which provides continuous coverage over 3200--8000 \AA{} with a dispersion of 1.96 \AA{}/pixel.
We utilized HeAr/Ne lamp exposures for wavelength calibration.
To obtain relative flux calibration sufficient for the measurement of line ratios, we applied a sensitivity curve derived from observations of the standard star G24-9.
All observations were conducted at parallactic to minimize slit losses.
The data were reduced using a custom \texttt{python} reduction pipeline, which found arc solutions and performed overscan bias correction and flat-fielding, median-combination of the individual data frames, rectification, background subtraction, and airmass-corrected flux normalization using the standard star observations.

\begin{table}
	\centering
	\caption{Summary of MMT observations on the night of January 25, 2017.}
	\label{tab:mmt_obs}
	\begin{tabular}{lcc}
		\hline
        Name & Airmass & Integration time (s)\\
		\hline
        \multicolumn{3}{c}{January 25, 2017} \\
        \hline
        HS1442+4250 & 1.1 & 6300 \\
        J0940+2935 & 1.3 & 7200 \\
        J1119+5130 & 1.2 & 4500 \\
        SBSG1129+576 & 1.2 & 5400 \\
        UM133 & 1.1 & 6300 \\
		\hline
	\end{tabular}
\end{table}

We also observed the five northern targets with ESI at Keck II in January and February 2017.
The exposure times are summarized in Table~\ref{tab:esi_obs}, and average to 2 hours per target.
Conditions were generally good, with typical seeing of 0.8\arcsec{}, 1.0\arcsec{}, and 1.1\arcsec{} on January 20, 21, and February 21 2017 (respectively).
We utilized the 1\arcsec{}x20\arcsec{} slit, producing spectra spanning 3900--10900 \AA{} with 11.5 km/s/pixel dispersion.
The standard set of HgNeXeCuAr arcs, dome flats, and biases were conducted at the beginning of the night, and archival sensitivity corrections were used to provide relative flux calibration of our observations (we only report line ratios from these data).
The continuum was sufficiently bright to trace the object through the echelle orders directly, and the spectra were extracted with a boxcar ensuring all object flux was captured.
We reduced these data with the \texttt{ESIRedux} code\footnote{\url{http://www2.keck.hawaii.edu/inst/esi/ESIRedux/}}.

\begin{table}
	\centering
	\caption{Summary of ESI observations of our XMPs conducted in January--February 2017.}
	\label{tab:esi_obs}
	\begin{tabular}{lcc}
		\hline
        Name & Airmass & Exposure (s)\\
		\hline
        \multicolumn{3}{c}{January 20 2017} \\
        \hline
        UM133 & 1.3 & 9000 \\
		\hline
        \multicolumn{3}{c}{January 21 2017} \\
        \hline
        UM133 & 1.2 & 1800 \\
		\hline
        \multicolumn{3}{c}{February 21 2017} \\
        \hline
        J0940+2935 & 1.2 & 7200 \\
        J1119+5130 & 1.2 & 4800 \\
        SBSG1129+576 & 1.2 & 9000 \\
        HS1442+4250 & 1.1 & 7200 \\
		\hline
	\end{tabular}
\end{table}

One target (J0405-3648) resides at declinations inaccessible to MMT or Keck.
For this object, we rely on optical line measurements and metallicity found in a consistent manner for the main body of the galaxy from a VLT/FORS2 spectrum by \citet{Guseva2009}.

The strong optical lines constrain the physical conditions and composition of the ionized gas in these star-forming regions.
We measure emission line fluxes and rest-frame equivalent widths in the optical spectra as described in \citetalias{Senchyna2017}, utilizing the Markov chain Monte Carlo sampler \texttt{emcee} \citep{Foreman-Mackey2013} to infer uncertainties in these measurements.
To correct for reddening in line measurements, we perform a two-step process, correcting first for Galactic extinction using the maps of \citet{Schlafly2011} with the $R_\text{V}=3.1$ extinction curve of \citet{Fitzpatrick1999}.
We then estimate the residual intrinsic reddening using the Balmer decrement relative to the Case B recombination value of H$\alpha/$H$\beta = 2.86$ \citep[assuming $T_e=10^4$ K, $n_e=10^3$ \unit{cm^{-3}}; ][]{Draine2011}.
We utilize the SMC bar average extinction curve measured by \citet{Gordon2003} for modeling the intrinsic attenuation.

The corrected optical nebular lines reveal a range of gas conditions prevail in these XMPs.
The \ott{} $=$ [\oiii{}] $\lambda 4959$ + $\lambda 5007$ / [\oii{}] $\lambda \lambda 3727, 3729$ ratios (Table~\ref{tab:basicopt}), which correlates with the density of ionizing radiation in \hii{} regions, varies from 0.6 in J0940+2935 (close to typical for local star-forming galaxies in SDSS) to 10.3 in HS1442+4250, near the largest values observed in the most intense star-forming regions \citep[e.g.][]{Senchyna2017,Izotov2017}.
The equivalent width of H$\beta$, which correlates with specific star formation rate and the degree to which the youngest stars dominate the optical, ranges from 18--94 \AA{}, corresponding to sSFRs assuming constant star formation of 1--20 \unit{Gyr^{-1}}, similar to the range found in Section~\ref{sec:imaging}. 

\begin{table*}
    \centering
    \caption{Optical line measurements from the MMT spectra, including metallicities computed using the direct method from [\oiii{}] $\lambda 4363$ detections. We also include measurements from a VLT/FORS2 spectrum of J0405-3648 presented by \citet{Guseva2009}, which we mark with a \textdagger. The final column presents the reddening-corrected ratio of \heii{} $\lambda 4686$/H$\beta$ measured from the ESI spectra (see Section~\ref{sec:heiixmp} for discussion).}
    \label{tab:basicopt}
\begin{tabular}{lcccccccc}
\hline
Name & \ott{} & \rtt{} & [\oiii{}] $\lambda 5007$ & H$\beta$ & E(B-V) & $T_e(\mathrm{\oiii{}})$ & $12+\log_{10}(\mathrm{O/H})$ & \heii{} $\lambda 4686$ /H$\beta$ $\times 100$\\ 
  &  &  & $W_0$ (\AA{}) & $W_0$ (\AA{}) &  & $10^4$ K &  & ESI\\ 
\hline
HS1442+4250& $10.1_{-0.7}^{+0.6}$& $7.1_{-0.2}^{+0.1}$& $515_{-33}^{+32}$& $113_{-5}^{+4}$& 0.01 & $1.70 \pm 0.07$ & $7.65 \pm 0.04$ & $3.6 \pm 0.1$\\ 
J0405-3648 \textdagger & 2.1 \textdagger & 4.1 \textdagger & -- & 18 \textdagger & -- & $1.50\pm 0.13$ \textdagger & $7.56\pm 0.07$ \textdagger & -- \\
J0940+2935& $0.7_{-0.0}^{+0.0}$& $5.6_{-0.2}^{+0.2}$& $46_{-2}^{+2}$& $26_{-1}^{+1}$& 0.07 & $1.65 \pm 0.30$ & $7.63 \pm 0.14$ & $< 1.2$\\ 
J1119+5130& $2.9_{-0.2}^{+0.2}$& $5.4_{-0.2}^{+0.2}$& $110_{-8}^{+7}$& $36_{-2}^{+2}$& 0.10 & $1.80 \pm 0.14$ & $7.51 \pm 0.07$ & $2.6 \pm 0.3$\\ 
SBSG1129+576& $1.6_{-0.1}^{+0.1}$& $4.6_{-0.3}^{+0.1}$& $145_{-4}^{+4}$& $69_{-2}^{+2}$& 0.09 & $1.79 \pm 0.14$ & $7.47 \pm 0.06$ & --\\ 
UM133& $2.9_{-27.5}^{+0.3}$& $6.4_{-1.7}^{+0.4}$& $301_{-36}^{+37}$& $85_{-8}^{+6}$& 0.10 & $1.58 \pm 0.10$ & $7.70 \pm 0.09$ & $2.1 \pm 0.1$\\ 
\hline
\end{tabular}
\end{table*}

With these de-reddened optical line measurements, we can measure the gas-phase oxygen abundance using a direct temperature estimate.
We follow the procedure outlined in \citet{Senchyna2019} using the calibrations of \citet{Izotov2006} to measure direct-$T_e$ oxygen abundances with the $T_e$-sensitive line [\oiii{}] $\lambda 4363$ and [\oii{}] $\lambda \lambda 3727,3729$ measured in the MMT spectra.
The results are displayed in Table~\ref{tab:basicopt}, and confirm the metal-poor nature of these systems.
The gas-phase abundances range from $12+\log\mathrm{O/H}=7.38$ -- 7.70, corresponding to $Z/Z_\odot \sim $ 0.05--0.1 assuming solar abundances.
Thus, our observations probe young stellar populations at metallicities approaching the lowest yet observed in star-forming galaxies, $Z/Z_\odot \simeq 0.02$ \citep[][and references therein]{Izotov2018}.
Moderate-resolution ultraviolet spectral constraints in this metallicity regime are nearly entirely absent (Section~\ref{sec:uvarchive}).

\subsection{UV Spectra}
\label{sec:uvspec}

\begin{figure*}
    \includegraphics[width=0.7\textwidth]{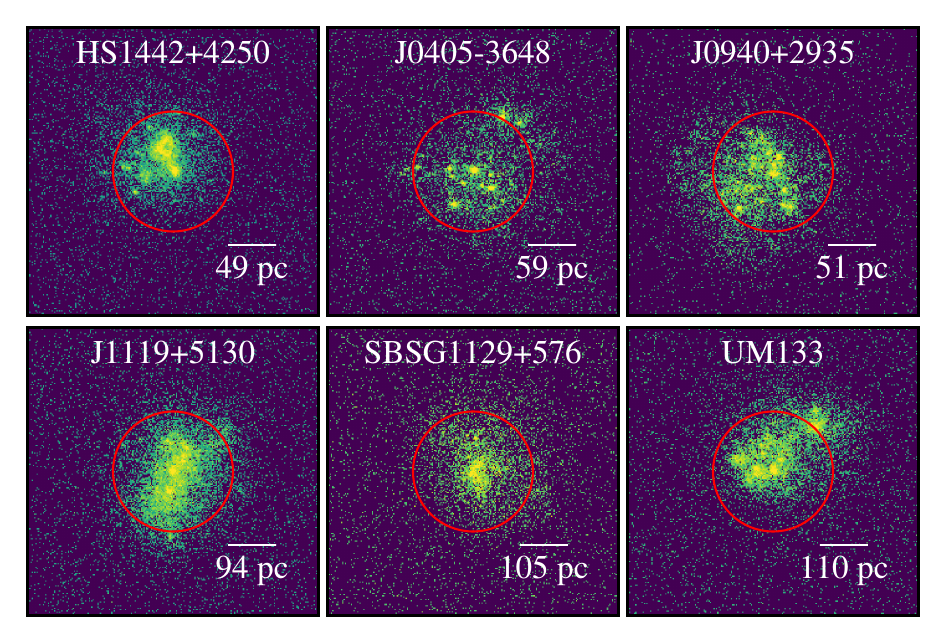}
    \caption{
        \hstcos{} NUV target acquisition images for the 6 XMPs.
        The red circle denotes the 2.5\arcsec{} COS spectroscopic aperture, and a 1\arcsec{} scalebar is presented with corresponding comoving distance estimates at the redshift of each system.
        At this spatial resolution, each compact object is resolved into multiple star-forming regions within the spectroscopic aperture.
    }
    \label{fig:tacq}
\end{figure*}

We obtained ultraviolet spectra with \hstcos{} in Cycle 24 (GO: 14168, PI: Stark) targeting the key nebular and stellar features at \civ{} $\lambda \lambda 1548, 1550$, \heii{} $\lambda 1640$, \oiii{}] $\lambda \lambda 1661,1666$, and \ciii{}] $\lambda \lambda 1907, 1909$.
The required spectral resolution to fully disentangle these features and interstellar absorption ($R\gg 1000$ after binning over several resolution elements; e.g.\ \citealt{Crowther2006a,Senchyna2017,Vidal-Garcia2017}) necessitated observations with the G160M and G185M grisms coupled to the 2.5\arcsec{} diameter Primary Science Aperture (PSA).
Our targets were selected to be bright enough for high S/N continuum observation with one orbit per grism, including an 88 second acquisition image taken in the NUV with MIRRORA.
An unexpected HST safing during observation of J0405-3648 resulted in an automatic re-observation, requiring two G185M exposures of this object.
After confirming that both achieved comparable signal-to-noise, we combined these two spectra for our analysis.
We plot these acquisition images in Figure~\ref{fig:tacq}, which reveal in very high-resolution the structure of the star-forming regions within the aperture.
While centrally-concentrated, these images indicate that we are observing the conglomerate spectra of several clusters spread over a $\sim 50$--100 pc region.

We used the TIME-TAG observation mode with all wavelength offset settings (FP-POS=ALL) and a wavelength calibration lamp exposure (FLASH=YES), minimizing the effects of fixed-patten noise in the detectors.
The data were reduced with \texttt{CALCOS} 3.2.1 (2017-04-28) and the default calibration files as of 2017-08-11.
The spectra were extracted with the default parameters, namely with the TWOZONE algorithm in the FUV and BOXCAR in the NUV.
The reduced spectra have dispersions of 73.4 m\AA{}/resolution element in G160M/FUV and 102 m\AA{}/resel in G185M/NUV.
We bin the one-dimensional specta via boxcar averaging over the length of these resolution elements (or multiples thereof) to achieve higher S/N per pixel for plotting and analysis.
Typical binning over 3 resolution elements \citep[as motivated by the measured width of MW absorption lines;][]{Senchyna2017} result in an effective spectral resolution of 40--50 km/s.

\begin{figure*}
    \includegraphics[width=1.0\textwidth]{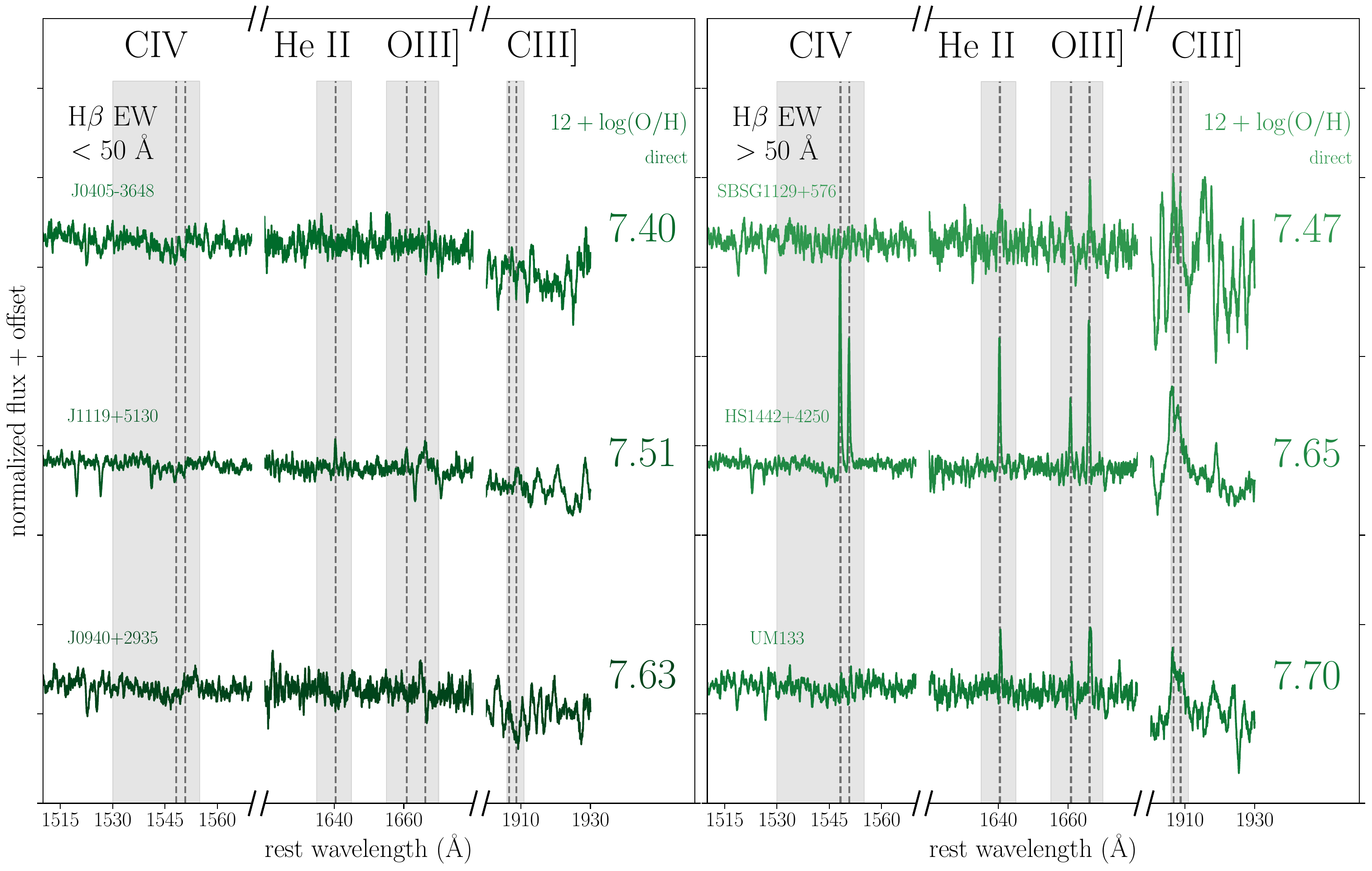}
    \caption{
        \hstcos{} UV spectra of the 6 XMPs targeted in this program.
        We plot the three systems with H$\beta$ equivalent widths $<50$ \AA{} in the left panel, and those with higher equivalent width optical emission on the right, ordered within each panel by gas-phase oxygen abundance.
        All spectra are normalized by their flux at 1650 \AA{} (the tick labels represent a separation of 2 in these normalized units) and separated by an arbitrary vertical offset for display.
        Despite their low gas-phase metallicity and correspondingly weak stellar winds, stellar \civ{} P-Cygni absorption is visible in most of the objects.
        The systems with high specific star formation rate and correspondingly high equivalent-width optical emission reveal typically more prominent UV nebular emission, with emission in nebular \civ{}, \heii{}, \oiii{}], and \ciii{}] detected.
    }
    \label{fig:majorlines_comp}
\end{figure*}

The spectra of our targets reveal a broad range of features, which we discuss on an object-by-object basis in Section~\ref{sec:uvobjectbyobject}.
We present line flux and equivalent width measurements, performed as described in \citetalias{Senchyna2017}, in Tables~\ref{tab:uvflux} and \ref{tab:uvew}, respectively.
We also plot the spectra in Figure~\ref{fig:majorlines_comp}, separated into two panels representing galaxies with H$\beta$ equivalent widths $>50$ \AA{} (left) and $<50$ \AA{} (right).
Nebular emission in \oiii{}] $\lambda 1666$ is detected in all galaxies except J0405-3648, and P-Cygni stellar absorption in \civ{} $\lambda \lambda 1548, 1550$ is visible in all but SBSG1129+576.

\begin{table*}
    \caption{UV nebular emission line flux measurements, with 3$\sigma$ upper limits where undetected.\label{tab:uvflux}}
\begin{tabular}{lcccccc}
\hline
Name & \civ{} 1548 & \civ{} 1550 & \heii{} 1640 & \oiii{}] 1661 & \oiii{}] 1666 & \ciii{}] 1907, 1909\\
 & ($10^{-15}$ ergs/s/cm$^2$)& ($10^{-15}$ ergs/s/cm$^2$)& ($10^{-15}$ ergs/s/cm$^2$)& ($10^{-15}$ ergs/s/cm$^2$)& ($10^{-15}$ ergs/s/cm$^2$)& ($10^{-15}$ ergs/s/cm$^2$)\\
\hline
\hline
HS1442+4250 & $4.52 \pm 0.05$ & $2.46 \pm 0.05$ & $2.64 \pm 0.05$ & $1.51 \pm 0.05$ & $3.26 \pm 0.05$ & $11.58 \pm 0.57$\\
J0405-3648 & <$0.06$ & <$0.06$ & <$0.07$ & <$0.22$ & <$0.22$ & <$0.51$\\
J0940+2935 & <$0.06$ & <$0.07$ & <$0.08$ & <$0.27$ & $0.69 \pm 0.04$ & <$0.57$\\
J1119+5130 & <$0.08$ & <$0.08$ & <$0.10$ & $0.90 \pm 0.06$ & $2.64 \pm 0.13$ & <$0.78$\\
SBSG1129+576 & <$0.04$ & <$0.05$ & $0.61 \pm 0.05$ & <$0.20$ & $0.42 \pm 0.02$ & <$0.76$\\
UM133 & <$0.09$ & <$0.09$ & $0.64 \pm 0.03$ & <$0.33$ & $1.44 \pm 0.04$ & $4.70 \pm 0.38$\\
\hline
\end{tabular}
\end{table*}

\begin{table*}
\begin{tabular}{lcccccc}
\hline
Name & \civ{} 1548 & \civ{} 1550 & \heii{} 1640 & \oiii{}] 1661 & \oiii{}] 1666 & \ciii{}] 1907, 1909\\
 & (\AA{})& (\AA{})& (\AA{})& (\AA{})& (\AA{})& (\AA{})\\
\hline
\hline
HS1442+4250 & $2.92 \pm 0.05$ & $1.46 \pm 0.03$ & $1.69 \pm 0.03$ & $0.99 \pm 0.03$ & $2.15 \pm 0.04$ & $11.09 \pm 0.72$\\
J0405-3648 & $<0.07$ & $<0.07$ & $<0.09$ & $<0.28$ & $<0.31$ & $<2.64$\\
J0940+2935 & $<0.07$ & $<0.06$ & $<0.08$ & $<0.27$ & $0.72 \pm 0.05$ & $<1.09$\\
J1119+5130 & $<0.04$ & $<0.04$ & $<0.05$ & $0.49 \pm 0.04$ & $1.53 \pm 0.08$ & $<0.83$\\
SBSG1129+576 & $<0.08$ & $<0.08$ & $1.14 \pm 0.09$ & $<0.41$ & $0.95 \pm 0.05$ & $<3.51$\\
UM133 & $<0.09$ & $<0.08$ & $0.60 \pm 0.03$ & $<0.32$ & $1.47 \pm 0.05$ & $10.75 \pm 1.13$\\
\hline
\end{tabular}
    \caption{Rest-frame equivalent width measurements for the UV nebular emission lines, with 3$\sigma$ upper limits where undetected.}
    \label{tab:uvew}
\end{table*}

\section{Results}
\label{sec:results}

In this section we present and analyze the new \hstcos{} UV spectra collected for six XMPs in \hst{} program GO:14679.
First, we describe the spectra of each object individually in the context of their optical data in Section~\ref{sec:uvobjectbyobject}.
Next, we investigate the range of UV nebular line equivalent widths and line ratios attained by galaxies in this metallicity range in Section~\ref{sec:uvnebprop}.
Finally, we explore \heii{} emission in these XMPs in Section~\ref{sec:heiixmp}, and the stellar \civ{} P-Cygni feature in Section~\ref{sec:civpcygni}.

\subsection{New UV Spectra Object-by-Object}
\label{sec:uvobjectbyobject}

The \hstcos{} FUV and NUV data we obtained alongside deep optical integrations reveal a diversity of spectral properties in the XMPs targeted (Figure~\ref{fig:majorlines_comp}).
Here, we describe the UV spectra in the context of our other measurements for each object individually, before physically interpreting these observations in later sections.

{\bf HS1442+4250:}
This system is the optically-dominant component of a `cigar-like' extended light distribution (Figure~\ref{fig:cutouts}), and presents the most intense optical emission line properties in our sample.
The intense H$\beta$ emission (equivalent width $113\pm 5$ \AA{}) and high \ott{} ratio ($10.3\pm 0.6$) indicate this system harbors a very young stellar population and highly-ionized gas.
The UV spectra of this object are also extreme: we detect strong nebular emission in all targeted lines, with prominent emission in both \ciii{}] at $W_0=11.6\pm 0.6$ \AA{} and \civ{} at $W_0=4.38 \pm 0.07$ \AA{} equivalent width.
Narrow \heii{} at nearly the flux of \oiii{}] 1666 is also present.
The signal-to-noise in the continuum is sufficient to clearly detect a \civ{} P-Cygni profile as well; with the nebular emission and interstellar absorption subtracted-out by performing Gaussian fits to each component, the remaining broad stellar absorption presents an equivalent width of $-1.6$ \AA{}.

{\bf J0405-3648:}
The target region in this case is at the center of the optical galaxy, but is somewhat more extended than the other targets.
The measured stellar mass of this object at $10^{5.3\pm 0.2}$ $M_\odot$ is fairly typical for the sample, but the H$\beta$ equivalent width of 18 \AA{} measured by \citet{Guseva2009} suggests a significantly less dominant young stellar population in this system than the other targets (spanning 26--113 \AA{}, Table~\ref{tab:basicopt}).
The UV spectra show no detections of the target nebular lines (Figure~\ref{fig:majorlines_comp}), with 3$\sigma$ limiting equivalent widths of $<2.6$ \AA{} for \ciii{}] and $<0.3$ \AA{} for \oiii{}] 1666.
However, a \civ{} P-Cygni profile is clearly visible with an absorption $W_0=-1.9$ \AA{}, the deepest measured in the sample.

{\bf J0940+2935:}
This system is a $10^{4.9\pm 0.2} M_\odot$ star-forming complex with an extended tadpole-like tail.
The $26\pm 1$ \AA{} equivalent width of H$\beta$ in this system is also substantially smaller than in HS1442+4250 and closer to that of J0405-3648, indicative of a much less dominant population of massive stars.
The UV spectra of this object reveal weak nebular emission, with only \oiii{}] 1666 detected ($W_0=0.7$ \AA{}), but fairly prominent \civ{} P-Cygni absorption ($W_0=-0.9$ \AA{}).

{\bf J1119+5130:}
This system is relatively compact in SDSS imaging, with stellar mass $10^{6.4\pm 0.2} M_\odot$ in the spectroscopic aperture and an H$\beta$ equivalent width of $36\pm 2$ \AA{} just below the median for our sample.
The UV spectra reveal nebular emission in \oiii{}] 1666 at $W_0 = 1.5$ \AA{}, but non-detections of the other lines (with \ciii{}] $W_0<0.8$ \AA{}).
\civ{} P-Cygni absorption is barely detected in J1119+5130, at only $-0.2$ \AA{} equivalent width.

{\bf SBSG1129+576:}
The targeted star-forming region in this galaxy is situated at the end of an extended light distribution, and contains a relatively large stellar mass of $10^{6.1\pm 0.2} M_\odot$ with H$\beta$ equivalent width $69\pm 2$ \AA{}.
The \hstcos{} data reveal narrow \heii{} 1640 emission at $1.14\pm 0.09$ \AA{} and \oiii{}] 1666 at $0.95\pm 0.05$ \AA{}, but a nondetection of \ciii{}] ($<3.5$ \AA{}) and no detection of \civ{} P-Cygni absorption.

{\bf UM133:}
The targeted region in UM133 is a star-forming complex at the end of an extended tadpole structure (Figure~\ref{fig:cutouts}).
We infer a stellar mass of $10^{5.9\pm 0.2} M_\odot$ and  measure an H$\beta$ equivalent width of $85\pm 7$ \AA{} (the second largest in the sample), suggestive of a system particularly dominated by young massive stars.
The UV spectra of UM133 show prominent nebular emission in \heii{}, \oiii{}], and \ciii{}], with the \ciii{}] doublet measured at a very large equivalent width of $11\pm 1$ \AA{}.
\civ{} P-Cygni absorption is also visible, at $W_0=-1.3$ \AA{}.

\subsection{UV Nebular Properties of the New XMPs}
\label{sec:uvnebprop}

The moderate-resolution \hstcos{} observations described above in conjunction with recently-acquired data with the G140L grating (see Section~\ref{sec:sample}) allow us to investigate the production of UV nebular lines in this relatively unexplored metallicity regime.
In particular, understanding what governs the strength of the semi-forbidden \ciii{}] doublet at primordial metallicities is of critical interest given the growing body of detections of and constraints on this line at high redshift \citep[e.g.][and references therein]{Stark2014,Stark2015,Jaskot2016,Stark2017,Ding2017,Du2017,Maseda2017,Nakajima2018}.
Local observations have provided evidence for an inverse trend in \ciii{} equivalent width with metallicity, as strong $>5$ \AA{} \ciii{}] emerges only at $12+\log\mathrm{O/H}<8.4$ ($Z/Z_\odot \lesssim 0.5$) \citep[][\citetalias{Senchyna2017}]{Rigby2015}.
Among the sample presented in \citetalias{Senchyna2017}, we found that very prominent nebular \ciii{}] at equivalent widths $>10$ \AA{} is powered down to $12+\log\mathrm{O/H} \simeq 7.8$, and found some evidence that the maximum \ciii{}] equivalent width may even increase down to this limiting metallicity.
However, we expect to eventually encounter a turnover in this trend.
At sufficiently low metallicity, the harder stellar ionizing spectrum (leading to higher \civ{}/\ciii{}]) and the declining carbon abundance conspire to suppress \ciii{}], producing a peak in \ciii{}] equivalent widths with metallicity similar to that found for optical [\oiii{}] emission \citep[e.g.][]{Jaskot2016}.
Testing this with observations of XMPs is crucial to establishing an empirical baseline for correctly physically interpreting high equivalent width \ciii{}] at high redshift.

\begin{figure*}
    \includegraphics[width=2\columnwidth]{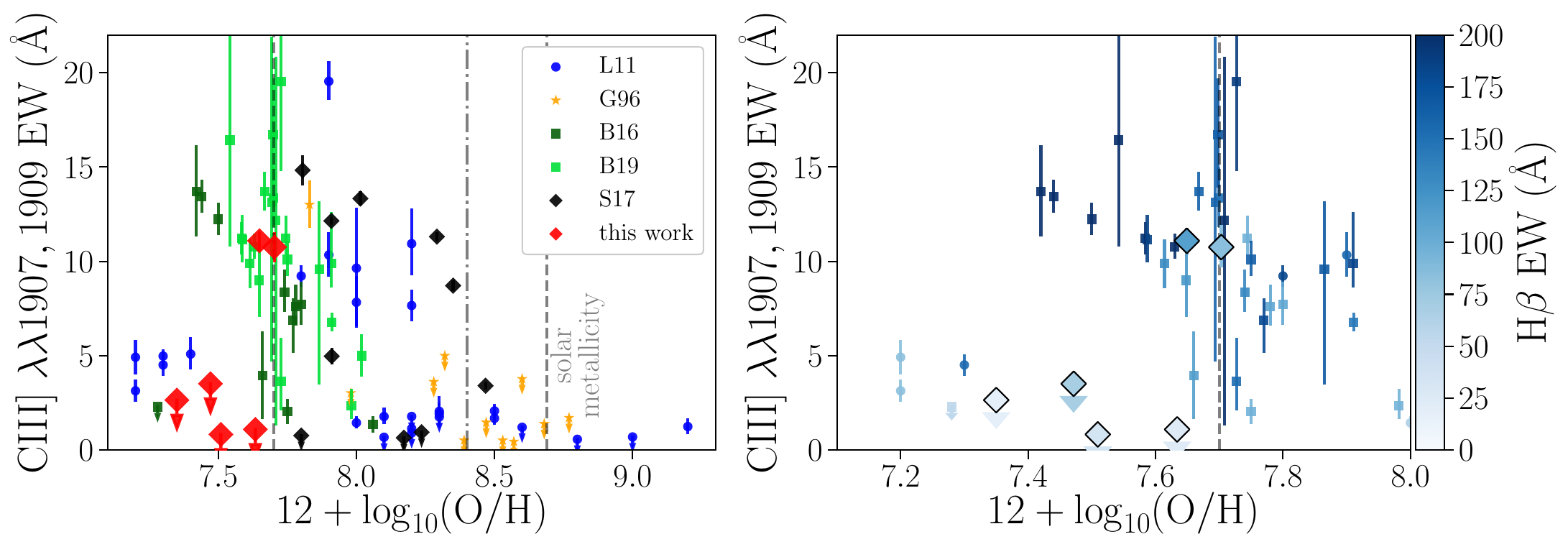}
    \caption{
        Equivalent width of the \ciii{}] $\lambda \lambda 1907, 1909$ doublet as a function of gas-phase metallicity.
        In addition to the \hstcos{} data presented in this paper and \citetalias{Senchyna2017}, we plot archival data from nearby galaxies cataloged by \citet[][L11]{Leitherer2011}, \citet[][G96]{Giavalisco1996}, and \citet[][B16]{Berg2016} --- see \citetalias{Senchyna2017} for more details.
        Our sample and that of \citet{Brorby2016} together demonstrate that strong $>10$ \AA{} \ciii{}] can be powered by very young stellar populations down to $12+\log\mathrm{O/H}\simeq 7.4$.
        The two XMPs presented here (HS1442+4250 and UM133) with strong 11 \AA{} \ciii{}] emission are also those with the highest equivalent width optical line emission (H$\beta$ equivalent widths $> 80$ \AA{}), corresponding to a very prominent young stellar population.
        In the right panel, we replot those systems at $12+\log\mathrm{O/H}<8.0$ with H$\beta$ equivalent width measurements from SDSS or similar \citep[][and Section~\ref{app:archive}]{Berg2016,Senchyna2017,Berg2019}.
        The systems with prominent \ciii{}] emission $W_0>10$ \AA{} also present high equivalent width H$\beta$ emission in-excess of 100 \AA{}, indicating that \ciii{}] requires a stellar population weighted towards very young stars even in the XMP regime.
    }
    \label{fig:metal_ciii}
\end{figure*}

Our sample of XMPs with new UV spectra includes two with very prominent \ciii{}] emission, HS1442+4250 and UM133, both at an equivalent width of 11 \AA{}.
Emission at this strength is rare among local star-forming galaxies at any metallicity, as illustrated in Figure~\ref{fig:metal_ciii}.
Intriguingly, the strong \ciii{}]-emitters are the highest metallicity systems in our new sample at $12+\log\mathrm{O/H}=7.65$ and 7.70.
The non-detection of \ciii{}] in our lower-metallicity systems is at first suggestive of a possible decline in \ciii{}] equivalent widths at $12+\log\mathrm{O/H}<7.6$.
However, the \ciii{}] emitters HS1442+4250 and UM133 also host the highest specific star formation rates of the systems we targeted.
Their characteristic stellar ages from SED fitting are very young at 45 and 47 Myr, and they present correspondingly large H$\beta$ equivalent widths ($W_0=94$ and 81 \AA{}, respectively).
Thus, the apparent decline in \ciii{}] among our lowest metallicity systems is likely due to their relative lack of ionizing flux from very young stellar populations.

Examining the full set of XMPs confirms the importance of specific star formation rate in determining the prominence of \ciii{}] even at very low metallicity.
In the right panel of Figure~\ref{fig:metal_ciii}, we replot the points at $12+\log\mathrm{O/H}<8.0$ color-coded by the equivalent width of H$\beta$ as measured in an SDSS spectrum or comparable (Section~\ref{app:archive}), which serves as a proxy for specific star formation rate.
Six XMPs at $12+\log\mathrm{O/H}<7.6$ from \citet{Berg2016,Berg2019} power H$\beta$ at equivalent widths $>200$ \AA{}, and all six present \ciii{}] emission at equivalent widths of 11--16 \AA{} comparable to the strongest measured at any metallicity in the local Universe (Figure~\ref{fig:metal_ciii}, right panel).
This suggests that we have not yet observed a significant turnover in the strength of \ciii{}] down to $12+\log\mathrm{O/H}\simeq 7.4$.
Rather, even at very low metallicities, \ciii{}] remains strong ($\gtrsim 10$ \AA{}) in systems characterized by young effective stellar ages ($<100$ Myr).

Nebular emission in \civ{} is also visible in the system with the most prominent young stellar population, HS1442+4250.
The \civ{} doublet is detected at a combined equivalent width of 4.4 \AA{}.
This is larger than any of the extreme star-forming regions at higher metallicities ($12+\log\mathrm{O/H}>7.7$) explored in \citetalias{Senchyna2017}, and is the fourth largest of any nearby star-forming system in the literature (see Section~\ref{sec:discuss}).
The total observed flux of the \civ{} doublet in HS1442+4250 is $0.6\times$ that in \ciii{}] even before correction for dust extinction, illustrating that these emission lines can be comparable in observed strength in local extremely metal-poor systems.
We will investigate the \civ{} doublet and its dependence on metallicity and sSFR in more detail in the discussion (Section~\ref{sec:discuss}).

\begin{figure}
    \includegraphics[width=\columnwidth]{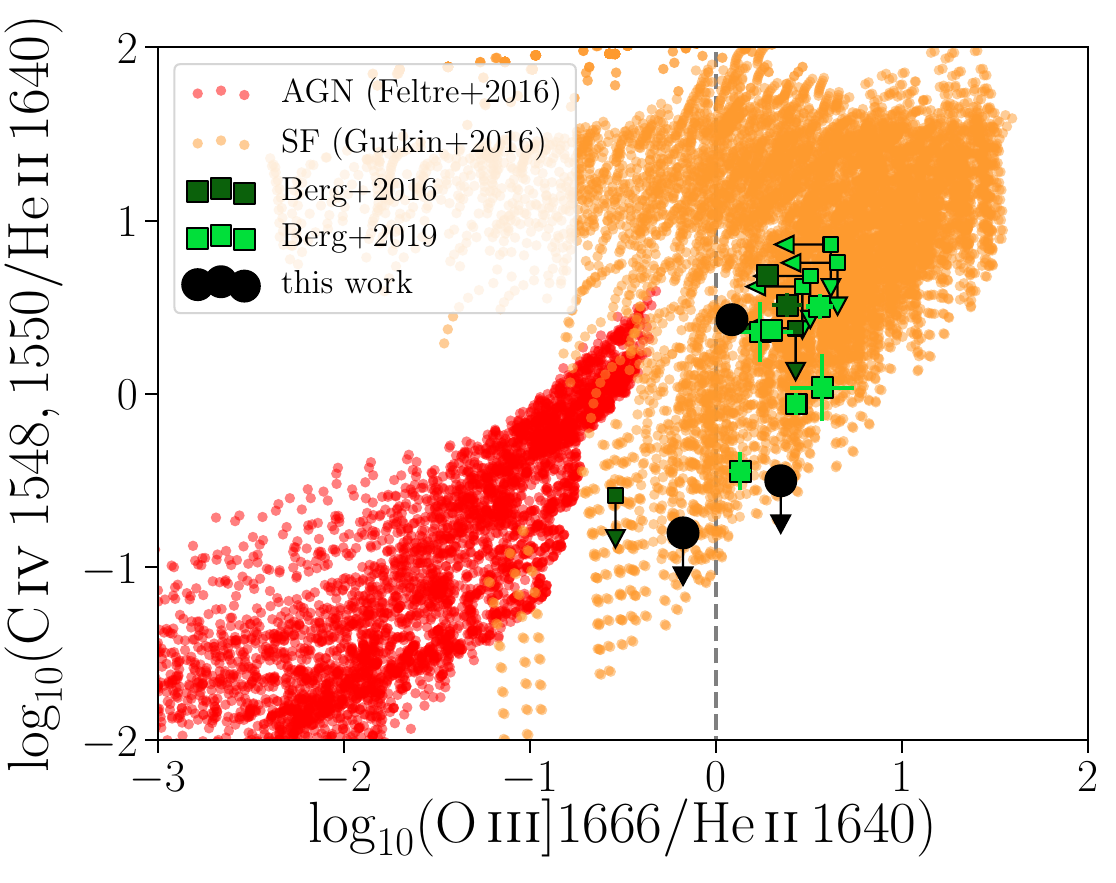}
    \caption{
        An ultraviolet line ratio diagnostic diagram, plotting de-reddened nebular \civ{} $\lambda \lambda 1548, 1550$ /\heii{} $\lambda 1640$ versus \oiii{}] $\lambda 1666$ / \heii{} $\lambda 1640$ for local star-forming XMPs (from spectra presented by this work and \citet{Berg2016,Berg2019}).
        Measurement uncertainties are plotted, but are often too small to be visible behind the points (of-order 0.01 dex).
        We also include the results of photoionization models of AGN from \citet{Feltre2016} and star-forming galaxies from \citet{Gutkin2016}.
        The new XMPs presented here remain disjoint from the nonthermal AGN models in this parameter space, but reside at the edge of the star-forming sequence as well in the \civ{}/\heii{} ratio.
        This suggests that the \heii{} or \civ{} emission in this metallicity regime is not fully captured by the existing models.
    }
    \label{fig:feltre_heii}
\end{figure}

Due to their sensitive dependence on the slope of the EUV ionizing spectrum, we might expect the ratios of the UV nebular lines to clearly distinguish stellar from nonthermal ionizing sources at high redshift \citep[e.g.][]{Feltre2016,Byler2018,Hirschmann2019}.
However, the realization that stellar models presently underpredict the strength of \heii{} emission in local star-forming galaxies and the resonant nature of \civ{} have raised doubts as to whether diagnostics involving these lines can be used, especially at low metallicities.
In Figure~\ref{fig:feltre_heii}, we plot the dust-corrected ratios of \oiii{}] $\lambda 1666$/\heii{} $\lambda 1640$ versus \civ{} $\lambda \lambda 1548,1550$ / \heii{} $\lambda 1640$ for all XMPs presented here and in \citet{Berg2016} and \citet{Berg2019} which have detections of the relevant lines (Section~\ref{app:archive}; we measure all lines and upper limits consistently using our line fitting code).
We overlay photoionization models powered by AGN and stellar spectra from \citet{Feltre2016} and \citet{Gutkin2016}, respectively.

The XMPs presented here are still entirely disjoint from the nonthermal models of \citet{Feltre2016} in this diagnostic diagram, residing close to the purely stellar model ratios.
All have dust-corrected \heii{} $\lambda 1640$ fluxes comparable to or smaller than \oiii{}] $\lambda 1666$, with only one system (SBSG1129+576) presenting slightly stronger \heii{} flux.
While the \oiii{}]/\heii{} ratios are only consistent with star-forming models, the \civ{}/\heii{} upper limits for SBSG1129+576 and UM133 place them at the bottom limit of the star-forming models presented.
This is suggestive of a $\mathrm{He}^{+}$-ionizing photon deficit in the stellar models considered, which we will discuss further in Section~\ref{sec:heiixmp}.
Despite this, the general disagreement of our local XMPs with the nonthermal models and similar early results at $z\sim 6$ \citep{Mainali2017} suggests that diagnostic diagrams such as this one still have utility in distinguishing star-forming and AGN systems even at very low metallicity, as long as the observations used are sufficiently high-resolution to disentangle nebular and stellar wind features.

\subsection{\heii{} Emission in the XMP regime}
\label{sec:heiixmp}

While the strength of \civ{} and \heii{} in these systems is inconsistent with a nonthermal power spectrum, the precise origin of the $\mathrm{He^+}$-ionizing photons remains unclear.
Though undisputed detections of nebular emission in \heii{} in predominantly star-forming systems at $z>6$ remain elusive \citep{Sobral2015,Mainali2017,Laporte2017,Shibuya2018,Sobral2019}, we expect this line to become very prominent in gas ionized by sufficiently low-metallicity stars \citep[e.g.][]{Tumlinson2000,Schaerer2003,Raiter2010}.
Previous observations have revealed that many low-metallicity star-forming galaxies produce these hard photons much more efficiently than predicted by stellar population synthesis models \citep[e.g.][]{Shirazi2012,Senchyna2017,Berg2018,Kehrig2018} which suggests significant systematic uncertainties may arise in the interpretation of \heii{} emission at high redshift.
In particular, we found in \citetalias{Senchyna2017} that the relative flux beyond the $\mathrm{He^+}$-ionizing edge is strongly dependent on metallicity, and suggests a dramatic upturn in the hardness of this ionizing radiation at low metallicities.
Our new data allow us to investigate whether this trend continues to lower metallicities, which may provide insight into the origin of this ionizing flux.

\begin{figure}
	\includegraphics[width=\columnwidth]{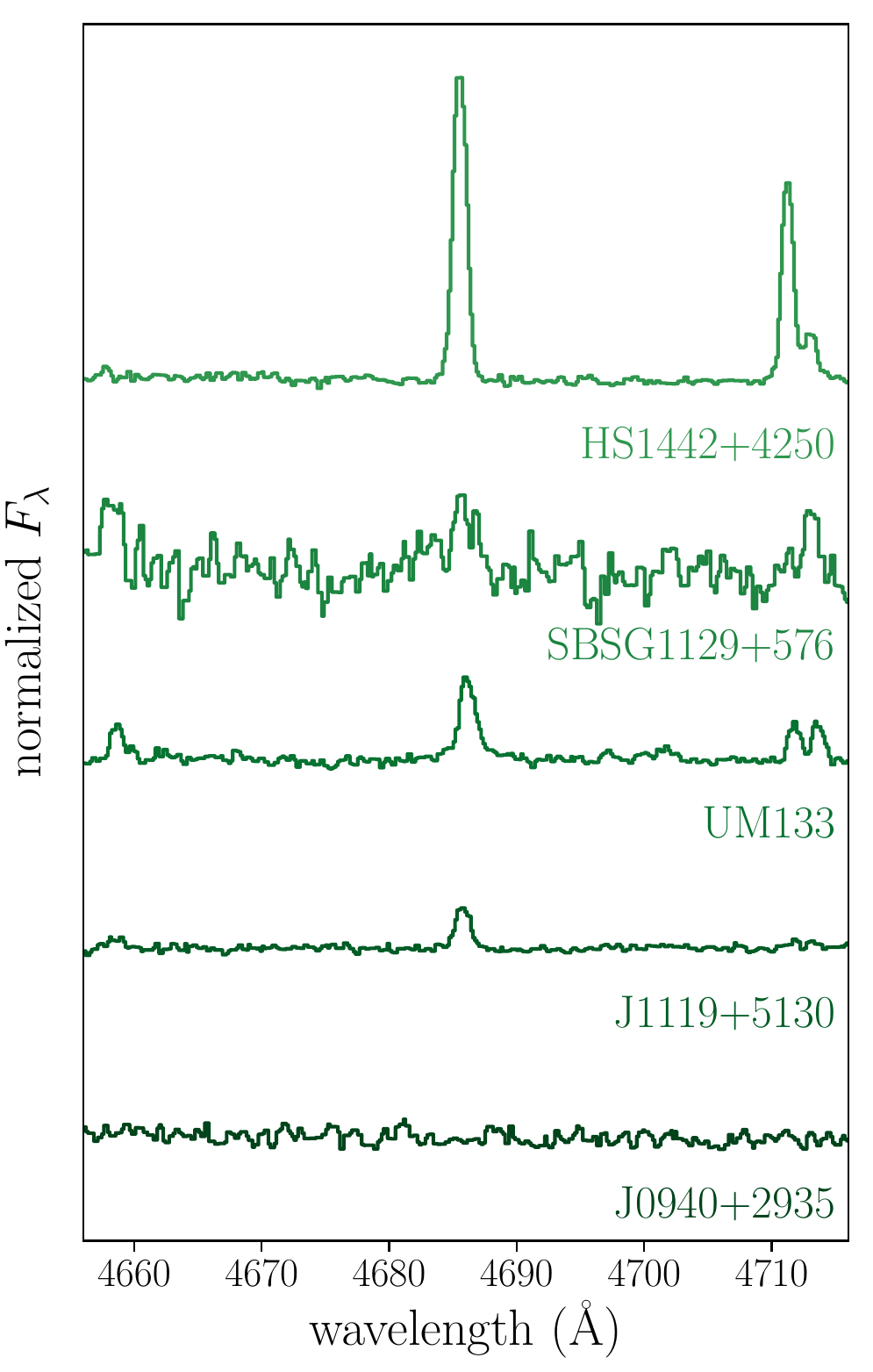}
    \caption{
        Cutouts of Keck/ESI optical spectra centred on the \heii{} $\lambda 4686$ line for the five northern hemisphere XMPs, ordered by the equivalent width of emission in this line.
        The \heii{} lines evident in HS1442+4250, UM133, and J1119+5130 are consistent in width with the other optical forbidden lines such as [\oiii{}], indicating that we are likely observing nebular emission from gas with kinematics similar to that which dominates these other lines.
    }
    \label{fig:esi_heii}
\end{figure}

We detect \heii{} $\lambda 1640$ in three of the six XMPs presented here, at equivalent widths up to 1.7 \AA{} in HS1442+4250 (Table~\ref{tab:uvew}).
The width of this \heii{} $\lambda 1640$ emission (as measured with uncertainties using our MCMC Gaussian fits without correction; Section~\ref{sec:uvspec}) is found to be as-narrow or narrower than that measured in [\oiii{}] $\lambda 1666$ for the strongest \heii{} emitters, HS1442+4250 and UM133, corresponding to FWHM$\simeq 80$ km/s.
Likewise, the four Keck/ESI detections of \heii{} $\lambda 4686$ (see Figure~\ref{fig:esi_heii}) reveal linewidths consistent with that measured for H$\beta$ to within $2\sigma$ in HS1442+4250, UM133, and J1119+5130.
In contrast, the \heii{} $\lambda 1640$ emission in SBSG1129+576 is fit by a single component with a substantially larger FWHM of $253 \pm21$ km/s, which indicates it likely arises in a stellar wind rather than ionized interstellar gas.
The linewidths measured for HS1442+4250, UM133, and J1119+5130 in the high-resolution data strongly suggests that \heii{} emission in these systems arises in gas similar to that which dominates the emission in nebular H$\beta$ and [\oiii{}], rather than a stellar wind or shocked outflow.

\begin{figure}
    \includegraphics[width=\columnwidth]{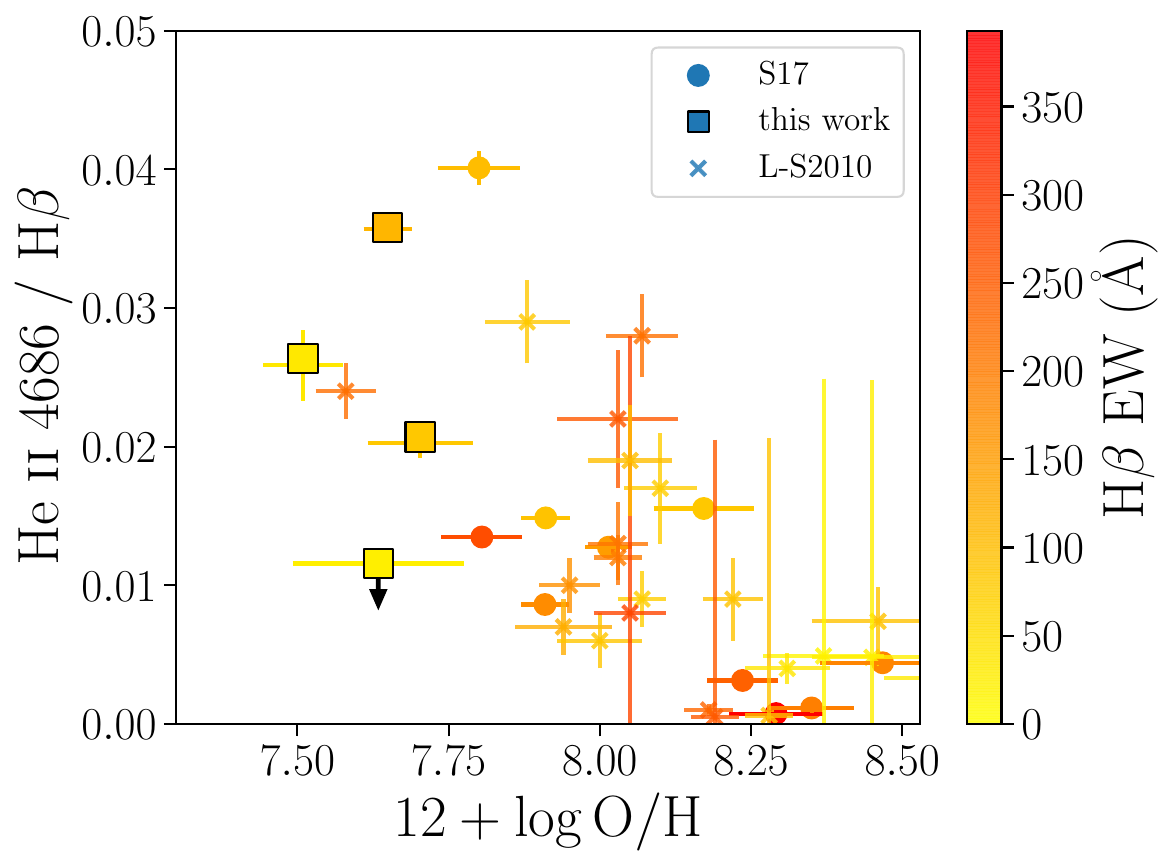}
    \caption{
        Optical \heii{}$\lambda 4686$ /H$\beta$ as a function of gas-phase oxygen abundance and H$\beta$ equivalent width, including the four galaxies with nebular detections or upper limits from ESI spectra from this sample and the ten from \citetalias{Senchyna2017}, as well as systems with high-resolution spectroscopy presented by \citet{Lopez-Sanchez2010}.
        Even at specific star formation rates far smaller than those of the sample presented by \citetalias{Senchyna2017} (H$\beta$ equivalent widths $\sim$ 20--100 \AA{} versus 90--400 \AA{}), we find large \heii{}/H$\beta$ ratios indicative of significant flux beyond $>54.4$ eV relative to H-ionizing flux in nearly all observed systems below $12+\log\mathrm{O/H}<8.0$ ($Z/Z_\odot<1/5$).
        This suggests that a metallicity-dependent source of hard EUV flux may be present on timescales longer than the lifetime of the most massive stars in these systems, though more data is needed to clarify this at the lowest metallicities $12+\log\mathrm{O/H}<7.7$.
    }
    \label{fig:heiihbeta}
\end{figure}

To assess the hardness of the ionizing radiation field powering \heii{} emission in these XMPs, we examine the flux ratio of \heii{} $\lambda 4686$ / H$\beta$ measured in the high-resolution ESI data.
This recombination line ratio tracks the ratio $Q(>54.4 \mathrm{eV})/Q(>13.6 \mathrm{eV})$, providing a relatively dust- and model-insensitive probe of the relative amount of hard ionizing flux beyond the $\mathrm{He^+}$-ionizing edge.
In \citetalias{Senchyna2017} we showed this ratio was elevated in systems below $12+\log\mathrm{O/H}<8.0$, suggestive of an origin in low-metallicity stars; but this sample was limited to metallicities $12+\log\mathrm{O/H}\gtrsim 7.8$.
In Figure~\ref{fig:heiihbeta}, we plot this flux ratio versus gas-phase metallicity for the four new XMPs with nebular \heii{} constraints alongside the galaxies presented in \citetalias{Senchyna2017} and those in \citet{Lopez-Sanchez2009,Lopez-Sanchez2010}.
The three XMPs in which we detect apparently nebular \heii{} emission reveal flux ratios \heii{}/H$\beta$ $=0.020$--$0.036$, indicating that the spectra of these systems are harder at the $\mathrm{He}^+$-ionizing edge than in all but one of the higher-metallicity extreme star-forming regions presented in \citetalias{Senchyna2017}.

Since the spread in metallicity among these three XMPs is small (0.2 dex), other parameters are likely at play in driving the large spread in \heii{}/H$\beta$.
If the flux beyond 54.4 eV were powered primarily by very massive stars, we would expect to see a correlation between this ratio and specific star formation rate and thus H$\beta$ equivalent width.
We have color-coded the measurements from all three samples in Figure~\ref{fig:heiihbeta} by \whb{}, which does not reveal a clear trend among the XMPs in this plot between this equivalent width and \heii{}/H$\beta$.
The XMP with the largest H$\beta$ equivalent width in this plot (SBS 1415+437 C with $\whb{}=222$ \AA{}, with measurements from \citealt{Lopez-Sanchez2009,Lopez-Sanchez2010}) presents a \heii{}/H$\beta$ ratio of only $0.024\pm0.002$, slightly smaller than that reached by J1119+5130 from our sample (\heii{}/H$\beta = 0.026\pm 0.003$) at a substantially smaller H$\beta$ equivalent width (36 \AA{}).
This suggests that while \heii{} emission is fairly ubiquitous at $12+\log\mathrm{O/H}<7.7$, sources with characteristic timescale longer than the lifetimes of the most massive stars may be partly responsible for powering it \citep[see also][]{Senchyna2019}.
Binary mass transfer can produce stars through both stripping and angular momentum transfer analogous to massive Wolf-Rayet stars with very hot emergent spectra but at a much broader range of initial masses and thus timescales \citep[e.g.][]{Eldridge2012,Goetberg2017,Goetberg2018}.
In addition, X-ray binaries may power substantial flux at the $\mathrm{He^+}$-ionizing edge \citep[e.g.][]{Fabbiano2006,Schaerer2019}, and fast-radiative shocks may also produce sufficient soft X-ray flux to contribute to \heii{} \citep[e.g.][]{Izotov2019}.
Distinguishing between possible sources will require a significantly expanded sample of well-measured line ratios on which models can be tested, especially at the lowest metallicities where this emission appears most common.

\subsection{The \civ{} Stellar P-Cygni Feature}
\label{sec:civpcygni}

Thus far, we have focused on the gas-phase oxygen abundance O/H inferred from the direct method in the optical.
However, the ionizing spectrum produced by stars is dependent primarily on Fe/H in these stars.
Variations away from fixed solar abundance ratios can lead to significant variations in the effective stellar metallicity at fixed O/H.
In particular, rising star formation histories can result in super-solar $\alpha/$Fe ratios, which has been invoked to explain the surprisingly low stellar metallicities inferred from spectra of star-forming systems at $z\sim 2$ \citep{Steidel2016,Strom2018}.
Thus, robustly connecting the inferences we draw about the ionizing spectrum from the observed nebular emission to stellar models will require us to consider carefully the metallicity of the stars present.
By providing access to metal features formed in the atmospheres of massive stars, ultraviolet spectra provide us leverage in assessing the actual stellar metallicity.
In particular, the stellar \civ{} P-Cygni feature is formed in the outflowing winds of massive O and B stars, and is visible in five of our six spectra (excepting SBSG1129+576; see Figure~\ref{fig:majorlines_comp}).

Since these stellar winds are assumed to be driven by the opacity of Fe-peak elements, the strength of this P-Cygni wind feature is highly correlated with the stellar metallicity \citep[e.g.][]{Crowther2006a}.
After subtracting nebular emission and interstellar absorption, the detected stellar absorption is found to vary in equivalent width from $-1.9$ to $-0.2$ \AA{}, with the two deepest features occurring in J0405-3648 (-1.9 \AA{}) and HS1442+4250 (-1.6 \AA{}).
As expected, the \civ{} P-Cygni feature is substantially weaker than the median equivalent width of -3.2 \AA{} measured in the sample of higher-metallicity star-forming regions in \citetalias{Senchyna2017}.
However, since the strength of this feature is sensitive to both metallicity and the exact mix of O and B stars present in the galaxy, we must consider both metallicity and stellar population age to derive stellar metallicity constraints from this feature.

\begin{figure}
    \includegraphics[width=\columnwidth]{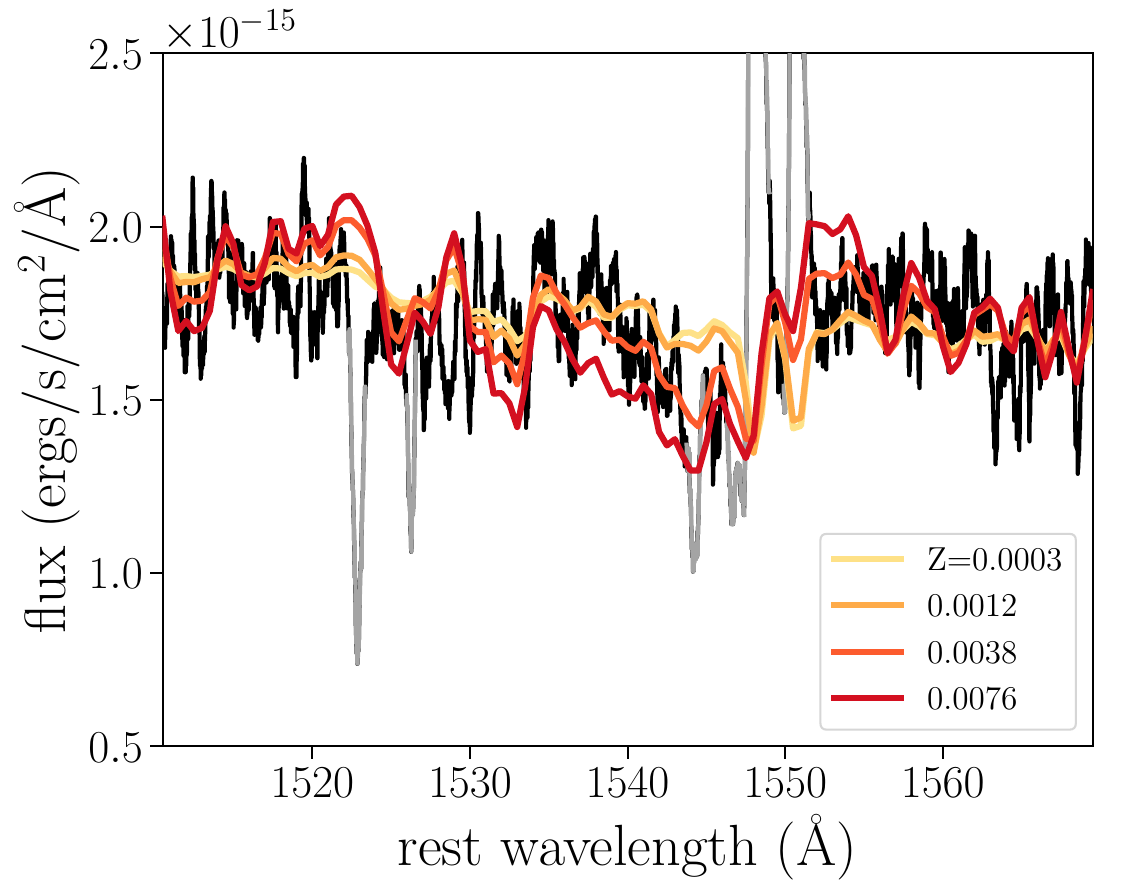}
    \caption{
        The \hstcos{} spectrum of HS1442+4250 centered on the \civ{} $\lambda \lambda 1548, 1550$ region.
        In addition to nebular emission, a broad stellar P-Cygni profile is clearly detected in this source.
        For clarity, we grey-out the portions of the spectrum impacted by nebular \civ{} emission and narrow absorption in \civ{} and \siII{} $\lambda 1526$ produced by both Milky Way gas and gas near the galaxy systemic redshift.
        We overlay purely-stellar continuum-normalized constant star formation history models produced by BEAGLE matched to the measured H$\beta$ equivalent width of this system at various metallicities.
        This comparison reveals that the massive stars in this system are inconsistent with the very low-metallicity models corresponding to super-solar $\alpha/$Fe ratios ($Z\sim 0.0003$--$0.001$), suggesting that the abundances are closer to solar than expected in systems at $z>2$.
    }
    \label{fig:civ_comp}
\end{figure}

The highest signal-to-noise detection of the P-Cygni feature is achieved in HS1442+4250, also the strongest nebular \civ{} emitter in our new sample of \hstcos{}-targeted XMPs.
We plot the G160M spectrum of this target focused on the stellar \civ{} profile in Figure~\ref{fig:civ_comp}, with the portions of the spectrum impacted by emission or absorption from gas greyed-out.
On top of this observed spectrum, we show continuum-normalized model stellar spectra computed assuming a constant star formation history with BEAGLE.
For each metallicity, we choose a stellar population age by matching the observed equivalent width of H$\beta$ to that predicted by the models.
Translating the gas-phase oxygen abundance of this system ($12+\log\mathrm{O/H}=7.66$) to $Z$ assuming the present-day solar metal abundances of \citet{Caffau2011} \citep[as in the opacity tables utilized by][]{Bressan2012} yields $Z=0.0012$. 
Uncertainties in the the scale of dust depletion in the gas-phase may increase the expected stellar metallicity, but only up to $Z\simeq 0.002$ \citep[see Table 2 of][]{Gutkin2016}.

If this system were as enhanced in $\alpha/$Fe as found at $z\sim 2$ (i.e. by a factor of 4 times), we would expect to find a stellar metallicity closes to $Z=0.0003$--$0.001$. 
However, the observed P-Cygni profile is substantially deeper than the shallow absorption predicted at this low metallicity.
In addition, the continuum around this wind line shows some indications of photospheric absorption in-line with that predicted by the higher-metallicity models but too faint to detect at $Z=0.0003$.
This suggests that this local \civ{}-emitter has an $\alpha/$Fe abundance much closer to solar than the enhanced values found at $z\sim 2$ and expected in the reionization era.
We will discuss the implications and caveats of this finding in more detail in Section~\ref{sec:discuss}.

\section{Discussion}
\label{sec:discuss}

Over the last several years, the body of reionization-era systems with rest-UV line detections has grown steadily with deep NIR spectroscopy campaigns.
These efforts have revealed strong \ciii{}] at equivalent widths up to 22 \AA{} and \civ{} at $\gtrsim 20$ \AA{} \citep[][]{Stark2015,Stark2017,Mainali2017,Schmidt2017}.
Early efforts to model these systems indicate that this prominent high-ionization emission may reflect the presence of very low metallicity stellar populations.
If confirmed, these modeling efforts suggest that the rest-UV might provide a means to identify extremely metal-poor systems in the reionization era.
However, stellar models suffer from substantial systematic uncertainties in the extremely metal-poor regime.
Since the Local Group contains no massive stars at $Z/Z_\odot<0.1$, direct calibration of temperature scales and winds cannot be conducted for individual stars at these metallicities.
Stellar models at $Z/Z_\odot<0.1$ are thus presently entirely theoretical, and predictions of the ionizing spectrum and the nebular emission it powers vary radically under different treatments of stellar evolution and radiation-driven winds \citep[e.g.][]{Kudritzki2002,Szecsi2015,Stanway2016}.
As a result, inferences about the nature of the stellar populations or other ionizing sources responsible for high-ionization emission in the reionization era are extremely uncertain.

Nearby dwarf galaxies provide a way to address these uncertainties in stellar population modeling at low metallicities.
Ultraviolet and optical spectra at low redshift allow us to investigate the variation of high-ionization lines with the metallicity and age of the underlying stellar population, providing empirical constraints on the ionizing spectra of young metal-poor stellar populations.
In \citet{Senchyna2017}, we presented moderate-resolution ultraviolet spectra of galaxies dominated by recent ($\lesssim 10$ Myr) star formation, spanning the metallicity range of $7.8<12+\log\mathrm{O/H}<8.5$ ($Z/Z_\odot \simeq 0.1$ -- $0.6$).
These data revealed nebular \civ{} only in the lowest metallicity systems $7.8<12+\log\mathrm{O/H}\lesssim 8.0$, with no detections even in the highest-sSFR systems at $12+\log\mathrm{O/H}\geq8.1$.
However, the equivalent width of \civ{} in this sample did not exceed 2 \AA{}, an order of magnitude lower than that measured at $z>6$ \citep{Stark2015a,Mainali2017,Schmidt2017}.
If the appearance of nebular \civ{} at $12+\log\mathrm{O/H}<8.0$ is due to a steepening of the stellar ionizing spectrum with decreasing metallicity, we might expect to encounter stronger emission in this doublet at $12+\log\mathrm{O/H}<7.7$.

Our new \hstcos{} spectra allow us to test whether more prominent \civ{} is powered in extremely metal-poor galaxies.
Indeed, we detect nebular \civ{} at a combined equivalent width of 4.4 \AA{} in HS1442+4250 ($12+\log\mathrm{O/H}=7.65$).
This \civ{} equivalent width is nearly three times larger than the maximum observed among the \citetalias{Senchyna2017} sample at $12+\log\mathrm{O/H}>7.7$.
In addition, the archival spectra we have collated on XMPs (Section~\ref{sec:uvarchive}) reveal three additional star-forming systems with \civ{} emission at equivalent widths in-excess of 4 \AA{} measured in lower-resolution \hstcos{} observations \citep[][]{Berg2016,Berg2019}, all at metallicities $12+\log\mathrm{O/H}<7.7$.
Together with nondetections at $12+\log\mathrm{O/H}>8.0$ \citep{Senchyna2017} these data suggest that nebular \civ{} is strongly metallicity dependent in star-forming galaxies, and reaches equivalent widths $>4$ \AA{} only in hot gas exposed to the hard stellar ionizing spectra produced at extremely low metallicities ($12+\log\mathrm{O/H}\lesssim 7.7$).

Given these results, we might expect to find \civ{} in most star-forming galaxies at sufficiently low metallicity.
Surprisingly, our data indicate that nebular \civ{} is far from ubiquitous even among XMPs.
One possibility is that strong \civ{} emission is powered only by very massive stars or other short-lived sources of ionizing radiation, in which case we would expect to see a trend with the effective age of the stellar populations.
To investigate this, we plot the equivalent width of nebular \civ{} versus \whb{} for the full sample of XMPs with \civ{} constraints assembled from our sample and the literature (see Appendix~\ref{app:archive}) in Figure~\ref{fig:civew_hbew}.
Nebular \civ{} is undetected in any XMPs with H$\beta$ equivalent widths $<100$ \AA{}, indicating that the ionization conditions required to power strong \civ{} are reached only for galaxies dominated by very young stellar populations ($<50$ Myr).
This full sample includes six systems with H$\beta$ equivalent widths $\gtrsim 200$ \AA{}, indicative of dominant star clusters of even younger characteristic age than in HS1442+4250 and closer to the most extreme galaxies studied in \citetalias{Senchyna2017} at higher metallicities.
All four of these galaxies are detected in \civ{} and one of the youngest, J104457, powers \civ{} emission at an equivalent width of 11 \AA{}.
This is by far the strongest \civ{} emission yet identified in local star-forming galaxies, and demonstrates that very young stellar populations at extremely low metallicity can power \civ{} at equivalent widths within a factor of several of that found at $z>6$.

\begin{figure}
    \includegraphics[width=\columnwidth]{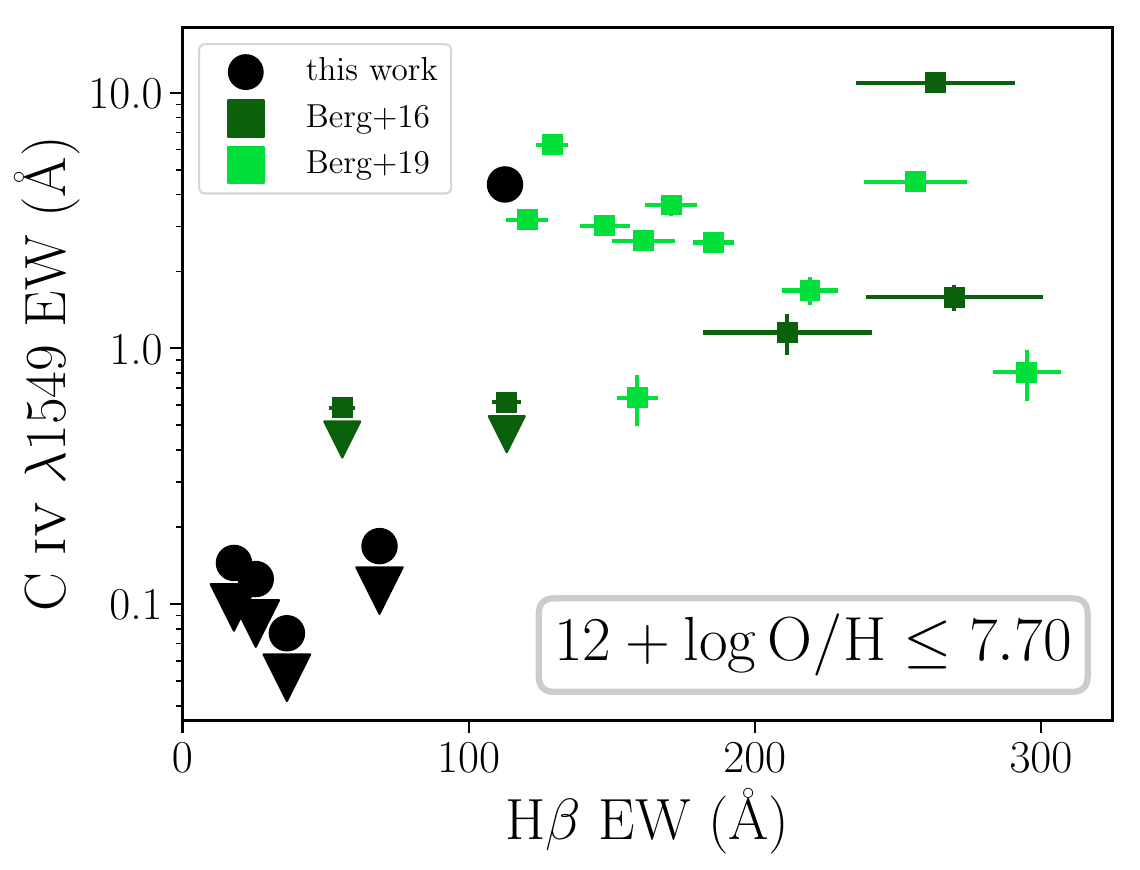}
    \caption{
        The equivalent width of nebular \civ{} $\lambda \lambda 1548,1550$ (plotted on a logarithmic scale) versus H$\beta$ equivalent width for XMPs with the necessary UV spectra.
        Nebular \civ{} emission is ubiquitous at metallicities $12+\log\mathrm{O/H}\leq 7.7$ among systems dominated by very young stars (H$\beta$ equivalent widths $>100$ \AA{}).
    }
    \label{fig:civew_hbew}
\end{figure}

While the strongest \civ{} arises from systems at $12+\log\mathrm{O/H}<7.7$ hosting very young stellar populations (characteristic ages $<50$ Myr), these conditions are not alone sufficient. 
Even among XMPs at H$\beta$ equivalent widths $\gtrsim 200$ \AA{}, we observe $\wciv{} \simeq 1$--11 \AA{} in systems with \civ{} constraints (Figure~\ref{fig:civew_hbew}).
This motivates a more detailed examination of the massive stars present in \civ{} emitters, whose actual composition and other properties may differ substantially from the simple picture inferred from the gas-phase oxygen abundance and \whb{} alone.
In particular, the galaxy $\alpha/$Fe ratio may be an important factor in determining the strength of \civ{}.
In systems which have very recently started forming stars rapidly, the buildup of iron and iron-peak elements in the ISM and stars is delayed relative to that of the $\alpha$ elements such as oxygen.
This is because type Ia supernovae are the primary source of iron enrichment, and occur on significantly larger timescales $>100$ Myr than the type II supernovae which dominate the production of $\alpha$ elements.
Since iron-peak elements are largely responsible for controlling the evolution and atmosphere opacities of massive stars, this elevated $\alpha$/Fe ratio results in a system with both a very hard low-metallicity stellar ionizing spectrum and significant quantities of oxygen and carbon to excite into nebular emission in the surrounding gas \citep[see][]{Steidel2016} --- likely ideal conditions for powering nebular \oiii{}], \ciii{}], and especially \civ{}.

Variation in $\alpha/$Fe and consequently the effective metallicity of the massive stars present could play a significant role in driving scatter in \wciv{} among high-sSFR XMPs.
Though the optical and UV spectra we have observed are dominated by stars formed recently, many lines of evidence indicate that even extremely metal-poor $z\sim 0$ star-forming dwarf galaxies have undergone multiple bursts of star formation over the past Gyrs \citep[e.g.][]{Tolstoy2009,Annibali2013,Janowiecki2017} which may lead some XMPs to values of $\alpha/$Fe closer to and potentially below solar \citep[e.g.][]{Bouret2015}.
Our comparison of the \civ{} P-Cygni profile in the spectrum of HS1442+4250 to stellar population synthesis models provides a first estimate of this ratio among local nebular \civ{} emitters (Section~\ref{sec:civpcygni}).
Indeed, the stellar winds in this system appear too strong to be consistent with the super-solar $\alpha/$Fe abundance ratios inferred at $z\sim 2$--3 \citep{Steidel2016,Strom2018}.

Completing a coherent picture of nebular \civ{} production will require additional measurements of $\alpha$/Fe and general massive star properties in other \civ{} emitters using moderate-resolution ultraviolet spectra.
Ideally, such spectra would be sufficiently deep to probe fainter photospheric absorption features which provide more direct estimates of $\mathrm{Fe/H}$ \citep[e.g.][]{Rix2004}.
If confirmed, a local trend between elevated $\alpha/$/Fe and enhanced \civ{} emission may also naturally explain the apparent rarity of $\gtrsim 20$ \AA{} \civ{} among local systems.
At high redshift where galaxies are increasingly characterized by rapidly rising star formation histories, we are likely to encounter very super-solar O/Fe ratios, as already demonstrated at $z\sim2$ \citep{Steidel2016}.
Extreme values are likely even more common at $z>6$ where the age of the Universe approaches the typical timescale for onset of Type Ia supernovae, and Pop III stellar yields become a significant consideration \citep[e.g.][]{Takahashi2014}.
While an accurate measurement of $\mathrm{Fe/H}$ in galaxies at reionization-era redshifts would be prohibitively expensive, strong wind lines such as \civ{} will likely be accessible for the brightest systems at $z>6$ with \jwst{}, enabling an independent check on this model.

While further local work is necessary, the existing UV data on local XMPs presented above already provide new context for reionization-era spectral observations.
One of the key objectives of \jwst{} and ground-based NIR spectroscopic campaigns targeting reionization-era systems is the characterization of the stellar populations and physical conditions in the earliest star-forming environments.
While prominent emission from ionized helium has long been suggested as a signpost of Population III star formation \citep[e.g.][]{Partridge1967,Tumlinson2000,Schaerer2003}, relatively little attention has been paid to the possibility of confirming the nonzero but likely very low metallicity of stars formed after this very first episode of enrichment.
Based upon this local sample and samples at higher metallicity (e.g. \citetalias{Senchyna2017}), detections of strong nebular \civ{} accompanied by UV line ratios inconsistent with a nonthermal spectrum appear to be a positive indicator of extremely metal-poor $\lesssim 0.1 Z_\odot$ stars and gas.
Indeed, the existing \civ{} detections at or in excess of an equivalent width of 20 \AA{} \citep{Stark2015a,Mainali2017} may be indicative of both very metal-poor stars and significantly super-solar $\alpha$/Fe abundances.
Nondetections of the high-ionization lines at $z>6$ are more difficult to diagnose, but larger samples and deeper spectra in the local Universe will provide insight into the full set of conditions which drive \civ{} strength at low metallicities.
Regardless, these data suggest that relatively inexpensive rest-UV emission-line spectroscopy with \jwst{} may provide significant insight into the early enrichment history of star-forming galaxies.

\section{Summary}
\label{sec:summary}

We presented new \hstcos{} UV spectroscopic observations of 6 star-forming galaxies in the poorly-sampled metallicity regime of XMPs.
These data provide a clear view of stellar and nebular features at the transitions \civ{} $\lambda \lambda 1548, 1550$, \heii{} $\lambda 1640$, \oiii{}] $\lambda \lambda 1661,1666$, and \ciii{}] $ \lambda \lambda 1907,1909$.
Along with all published archival UV spectra of XMPs, these data allow us to conduct the first systematic analysis of high-ionization nebular line production at $12+\log\mathrm{O/H}<7.7$ ($Z/Z_\odot \lesssim 0.1$), providing new insight into massive stellar populations and gas conditions encountered at these very low metallicities.
Our conclusions are as follows:

\begin{enumerate}
    \item 
    We detect \ciii{}] at equivalent widths in excess of 10 \AA{} in systems at $12+\log\mathrm{O/H}= 7.4$--7.7 (Figure~\ref{fig:metal_ciii}), confirming that this doublet can remain very prominent well into the extremely metal-poor regime.
    Even at these metallicities, our data suggests that strong \ciii{}] still requires very high specific star formation rates $\gtrsim 10\, \mathrm{Gyr^{-1}}$.
    \item We find strong nebular \heii{} emission in three of the studied XMPs, with \heii{}/H$\beta$ ratios indicative of relatively more flux at 54.4 eV than in all but one of the higher metallicity sources studied in \citetalias{Senchyna2017}.
    Despite this trend towards harder spectra at $12+\log\mathrm{O/H}<7.7$, we do not find a clear association between elevated \heii{} and \whb{} (Figure~\ref{fig:heiihbeta}).
    This suggests that some source with characteristic timescales longer than the age of the most massive stars may contribute to the flux beyond the $\mathrm{He^+}$-ionizing edge.
    Additional data is needed to clarify this picture and distinguish between possible origins of this excess hard ionizing flux.
    \item We find nebular \civ{} at an equivalent width of 4.4 \AA{} in the highest-sSFR system in our sample, nearly three times larger than the highest value found in star-forming systems at $12+\log\mathrm{O/H}>7.7$ in \citetalias{Senchyna2017}.
    Together with archival detections up to $\wciv{}=11$ \AA{} in galaxies observed at lower spectra resolution \citep{Berg2016,Berg2019}, these data confirm that strong \civ{} is clearly associated with both extremely low metallicity and very young stellar populations $<50$ Myr (Figure~\ref{fig:civew_hbew}).
    The body of UV spectral observations at $z\sim 0$ suggest that strong nebular \civ{} emission is a promising indicator of massive very metal-poor stars at reionization-era redshifts.
    \item While XMPs dominated by metal-poor massive stars appear capable of powering strong nebular \civ{}, the majority of such systems found locally do not produce emission above $4$ \AA{} in equivalent width, well below the first measurements of this doublet in systems at $z>6$.
    We suggest that this discrepancy could be explained by variations in the $\alpha/$Fe abundance ratio.
    Our moderate-resolution spectra reveal stellar features in the 4.4 \AA{} equivalent width nebular \civ{} emitter which are inconsistent with a super-solar $\alpha/$Fe ratio (Figure~\ref{fig:civ_comp}).
    Given the $\gg 100$ Myr timescales required for significant iron enrichment, assembling galaxies at $z>6$ may be significantly enhanced in $\alpha/$Fe.
    Such systems host both hot iron-poor stars which power hard ionizing spectra and gas already moderately enriched in carbon, likely enhancing the strength of nebular \civ{}.
    Larger samples of high quality local XMP spectra are required to test the impact of $\alpha/$Fe variations on high-ionization line emission.
\end{enumerate}

Clearly, additional observations of XMPs hosting very young stellar populations will be crucial for empirically calibrating stellar population synthesis models for the \jwst{} and ELT era.
While these systems are rare among bright XMPs uncovered by large spectroscopic survey programs like SDSS, techniques to find XMPs dominated by young massive stars have recently been developed \citep{Hsyu2018,Senchyna2019}, and may substantially expand the horizons of this avenue of study when applied to next-generation surveys such as LSST.
While model constraints will require statistical analysis of large samples of very well-understood nearby dwarf galaxies, such work may eventually provide a firm physical grounding for inferences about stellar populations in the reionization era.

\section*{Acknowledgements}

The authors thank the anonymous referee for their helpful report.
Based on observations made with the NASA/ESA Hubble Space Telescope, obtained from the data archive at the Space Telescope Science Institute.
Support for program \#14679 was provided by NASA through a grant from the Space Telescope Science Institute, which is operated by the Association of Universities for Research in Astronomy, Inc., under NASA contract NAS 5-26555.
Observations reported here were obtained at the MMT Observatory, a joint facility of the University of Arizona and the Smithsonian Institution.
Some of the data presented herein were obtained at the W.M. Keck Observatory, which is operated as a scientific partnership among the California Institute of Technology, the University of California and the National Aeronautics and Space Administration.
The Observatory was made possible by the generous financial support of the W.M. Keck Foundation.
The authors wish to recognize and acknowledge the very significant cultural role and reverence that the summit of Mauna Kea has always had within the indigenous Hawaiian community.
We are most fortunate to have the opportunity to conduct observations from this mountain.

This project used public archival data from the Dark Energy Survey (DES). Funding for the DES Projects has been provided by the U.S. Department of Energy, the U.S. National Science Foundation, the Ministry of Science and Education of Spain, the Science and Technology FacilitiesCouncil of the United Kingdom, the Higher Education Funding Council for England, the National Center for Supercomputing Applications at the University of Illinois at Urbana-Champaign, the Kavli Institute of Cosmological Physics at the University of Chicago, the Center for Cosmology and Astro-Particle Physics at the Ohio State University, the Mitchell Institute for Fundamental Physics and Astronomy at Texas A\&M University, Financiadora de Estudos e Projetos, Funda{\c c}{\~a}o Carlos Chagas Filho de Amparo {\`a} Pesquisa do Estado do Rio de Janeiro, Conselho Nacional de Desenvolvimento Cient{\'i}fico e Tecnol{\'o}gico and the Minist{\'e}rio da Ci{\^e}ncia, Tecnologia e Inova{\c c}{\~a}o, the Deutsche Forschungsgemeinschaft, and the Collaborating Institutions in the Dark Energy Survey.
The Collaborating Institutions are Argonne National Laboratory, the University of California at Santa Cruz, the University of Cambridge, Centro de Investigaciones Energ{\'e}ticas, Medioambientales y Tecnol{\'o}gicas-Madrid, the University of Chicago, University College London, the DES-Brazil Consortium, the University of Edinburgh, the Eidgen{\"o}ssische Technische Hochschule (ETH) Z{\"u}rich,  Fermi National Accelerator Laboratory, the University of Illinois at Urbana-Champaign, the Institut de Ci{\`e}ncies de l'Espai (IEEC/CSIC), the Institut de F{\'i}sica d'Altes Energies, Lawrence Berkeley National Laboratory, the Ludwig-Maximilians Universit{\"a}t M{\"u}nchen and the associated Excellence Cluster Universe, the University of Michigan, the National Optical Astronomy Observatory, the University of Nottingham, The Ohio State University, the OzDES Membership Consortium, the University of Pennsylvania, the University of Portsmouth, SLAC National Accelerator Laboratory, Stanford University, the University of Sussex, and Texas A\&M University.

Based in part on observations at Cerro Tololo Inter-American Observatory, National Optical Astronomy Observatory, which is operated by the Association of Universities for Research in Astronomy (AURA) under a cooperative agreement with the National Science Foundation.

DPS acknowledges support from the National Science Foundation through the grant AST-1410155.
JC, SC and AVG acknowledge support from the European Research Council (ERC) via an Advanced Grant under grant agreement no.\ 321323-NEOGAL.
AVG also acknowledges support from the ERC via an Advanced Grant under grant agreement no.\ 742719-MIST.

This research made use of Astropy, a community-developed core Python package for Astronomy \citep{AstropyCollaboration2013}; Matplotlib \citep{Hunter2007}; NumPy and SciPy \citep{Jones2001}; the SIMBAD database, operated at CDS, Strasbourg, France; NASA's Astrophysics Data System; and services or data provided by the NOAO Data Lab (NOAO is operated by the Association of Universities for Research in Astronomy (AURA), Inc.\ under a cooperative agreement with the National Science Foundation).

\bibliographystyle{mnras}
\bibliography{uvgals}

\appendix

\section{Archival UV Targets}
\label{app:archive}

In Table~\ref{tab:archival_data}, we have assembled UV line measurements for all XMPs we located in published samples with constraints on \ciii{}].
Prior to \hstcos{}, only four XMPs were observed with UV spectrographs covering \ciii{}]: I Zw 18 (with three separate pointings targeting different subregions), SBS 0335-052, SBS 1415+437, and Tol 1214-277.
These observations were detailed in the spectral atlas assembled by \citet{Leitherer2011}.
Unfortunately, these \fos{} and GHRS spectra did not cover \civ{}.
As in \citetalias{Senchyna2017}, we ignore the \ciii{}] emitter Tol 1214-277, as the continuum is undetected and results in a highly uncertain estimated \ciii{}] equivalent width.
We identified an additional fifteen systems at $12+\log\mathrm{O/H}<7.7$ observed with \hstcos{} G140L presented in \citet{Berg2016,Berg2019} with constraints on both \ciii{}] and \civ{}.
Gas-phase oxygen abundances are presented from \citet{Leitherer2011} and \citet{Berg2016,Berg2019}.

We measured \ciii{}] and \civ{} equivalent widths in these spectra using our spectral line-fitting tools described above.
However, note that at the lower resolution of G140L spectra ($R\sim 2000$), interstellar absorption and underlying stellar wind contributions cannot be fully cleaned from the spectra and may contaminate the nebular emission measurements.
We also cross-matched these spectra with SDSS as in \citetalias{Senchyna2017}, and include measurements of H$\beta$ equivalent widths where an SDSS spectrum was available.

\begin{table*}
    \centering
	\caption{
        UV and optical measurements of XMPs collated from the UV samples \citet[][L11]{Leitherer2011}, \citet[][B16]{Berg2016}, and \citet[][B19]{Berg2019}.
        Gas-phase oxygen abundances are drawn from the source reference, and H$\beta$ equivalent widths were measured uniformly from SDSS spectra of the regions where available.
        We present measured equivalent widths of both nebular \ciii{}] and \civ{} or corresponding 3$\sigma$ upper limits derived using our line fitting techniques, noting that these low-resolution spectra cannot be cleaned of underlying contamination from stellar or interstellar absorption near \civ{}.
    }
	\label{tab:archival_data}
\begin{tabular}{lcccccc}
\hline
    Name & Reference & Instrument & $12+\log\mathrm{O/H}$ & H$\beta$ & \ciii{}] & \civ{}\\
    &  & & from atlas & $W_0$ \AA{} (SDSS) & $W_0$ \AA{} & $W_0$ \AA{} (nebular)\\
\hline
    IZw18 & \citet[][L11]{Leitherer2011} & \fos{}/G190H& 7.20 & --- & $4.9 \pm 0.9$ & ---\\
    IZw18-NW HIIR & L11 & \fos{}/G190H& 7.20 & $84 \pm 2$ & $3.1 \pm 0.6$ & ---\\
    IZw18-SE HIIR & L11& \fos{}/G190H& 7.30 & $155 \pm 9$ & $4.5 \pm 0.6$ & ---\\
    SBS 0335-052 & L11& GHRS/G140L& 7.30& --- & $5.0 \pm 0.3$ & ---\\
    SBS 1415+437 & L11&\fos{}/G190H& 7.40& --- & $5.1 \pm 0.9$ & ---\\

    J082555 & \citet[][B16]{Berg2016} & \hstcos{}/G140L & 7.42 &  $270 \pm 31$ &  $13.7 \pm 2.4$ &  $1.6 \pm 0.2$ \\
    J085103 & B16 & \hstcos{}/G140L & 7.66 &  $113 \pm 5$ &  $3.9 \pm 2.3$ &  $<0.6$ \\
    J104457 & B16 & \hstcos{}/G140L & 7.44 &  $263 \pm 28$ &  $13.4 \pm 0.9$ &  $10.9 \pm 0.3$ \\
    J120122 & B16 & \hstcos{}/G140L & 7.50 &  $211 \pm 29$ &  $12.2 \pm 0.9$ &  $1.2 \pm 0.2$ \\
    J141454 & B16 & \hstcos{}/G140L & 7.28 &  $56 \pm 5$ &  $<2.3$ &  $<0.6$ \\

    J084236 & \citet[][B19]{Berg2019} & \hstcos{}/G140L & 7.61 &  $129 \pm 6$ &  $9.9 \pm 1.3$ &  $6.3 \pm 0.5$ \\
    J095430 & B19 & \hstcos{}/G140L & 7.70 &  $148 \pm 9$ &  $13.3 \pm 1.6$ &  $3.0 \pm 0.3$ \\
    J113116 & B19 & \hstcos{}/G140L & 7.65 &  $120 \pm 7$ &  $9.0 \pm 2.0$ &  $3.2 \pm 0.2$ \\
    J120202 & B19 & \hstcos{}/G140L & 7.63 &  $295 \pm 12$ &  $10.8 \pm 0.7$ &  $0.8 \pm 0.2$ \\
    J121402 & B19 & \hstcos{}/G140L & 7.67 &  $161 \pm 11$ &  $13.7 \pm 1.0$ &  $2.6 \pm 0.2$ \\
    J132347 & B19 & \hstcos{}/G140L & 7.58 &  $256 \pm 18$ &  $11.2 \pm 0.9$ &  $4.5 \pm 0.3$ \\
    J133126 & B19 & \hstcos{}/G140L & 7.69 &  $159 \pm 7$ &  $13.1 \pm 8.6$ &  $0.6 \pm 0.1$ \\
    J141851 & B19 & \hstcos{}/G140L & 7.54 &  $219 \pm 10$ &  $16.4 \pm 5.6$ &  $1.7 \pm 0.2$ \\
    J171236 & B19 & \hstcos{}/G140L & 7.70 &  $171 \pm 9$ &  $16.7 \pm 2.9$ &  $3.6 \pm 0.3$ \\
    J223831 & B19 & \hstcos{}/G140L & 7.59 &  $186 \pm 7$ &  $11.1 \pm 1.3$ &  $2.6 \pm 0.2$ \\

\hline
\end{tabular}
\end{table*}

\label{lastpage}

\end{document}